\documentclass[structabstract]{aa}
\usepackage{amsmath, hyperref}
\usepackage{graphicx}
\usepackage{subcaption}
\usepackage{caption}
\usepackage{pifont}
\usepackage{xcolor}
\usepackage{txfonts, longtable, dcolumn, verbatim}
\usepackage[T1]{fontenc}
\usepackage{natbib}
\bibpunct{(}{)}{;}{a}{}{,}
\usepackage{threeparttable}
\usepackage{array} 

\usepackage[squaren,textstyle]{SIunits}

\newcommand{\adam}{\texttt{ADAM}}
\newcommand{\mistral}{\texttt{Mistral}}

\begin{document}

   \title{The shape of (7)~Iris as evidence of an ancient large impact?  \thanks{Based on observations made with ESO Telescopes at the Paranal Observatory under programme ID 199.C-0074 (PI: P.~Vernazza) and 086.C-0785 (PI: B.~Carry).}
}

   \author{
          J.~Hanu{\v s}\inst{1}
           \and
          M.~Marsset\inst{2, 3}
               \and
          P.~Vernazza\inst{4}
              \and
          M.~Viikinkoski\inst{5}
              \and
          A.~Drouard\inst{4}
           \and
          M.~Bro\v{z}\inst{1}
           \and
          B.~Carry\inst{6}
           \and
          R.~Fetick\inst{4}
           \and
          F.~Marchis\inst{7}          
           \and
          L.~Jorda\inst{4}
           \and
          T.~Fusco\inst{4}
           \and
          M.~Birlan\inst{8}
           \and
          T.~Santana-Ros\inst{9}
           \and
          E.~Podlewska-Gaca\inst{9, 10}
           \and
          E.~Jehin\inst{11}
           \and
          M.~Ferrais\inst{11}
           \and
          J.~Grice\inst{6,12}
           \and
          P.~Bartczak\inst{9}
           \and
          J.~Berthier\inst{8}
           \and
          J.~Castillo-Rogez\inst{13}
           \and
          F.~Cipriani\inst{14}
           \and
          F.~Colas\inst{8}
           \and
          G.~Dudzi\'{n}ski\inst{9}
           \and
          C.~Dumas\inst{15}
          \and
          J.~\v{D}urech\inst{1}
           \and
          M.~Kaasalainen\inst{5}
           \and
          A.~Kryszczynska\inst{9}
           \and
          P.~Lamy\inst{4}
           \and
          H.~Le Coroller\inst{4}
           \and
          A.~Marciniak\inst{9}
           \and
          T.~Michalowski\inst{9}
           \and
          P.~Michel\inst{6}
           \and
          M.~Pajuelo\inst{8, 16}
           \and
          P.~Tanga\inst{6}
           \and
          F.~Vachier\inst{8}
           \and
          A.~Vigan\inst{4}
           \and
          O.~Witasse\inst{14}
           \and
          B.~Yang\inst{17}
}

   \institute{
            Institute of Astronomy, Charles University, Prague, V Hole\v sovi\v ck\'ach 2, CZ-18000, Prague 8, Czech Republic
             $^*$\email{hanus.home@gmail.com}
             \and
             Department of Earth, Atmospheric and Planetary Sciences, MIT, 77 Massachusetts Avenue, Cambridge, MA 02139, USA
             \and
            Astrophysics Research Centre, Queen's University Belfast, BT7 1NN, UK
         \and
            Aix Marseille Universit\'e, CNRS, LAM, Laboratoire d'Astrophysique de Marseille, Marseille, France
         \and
            Department of Mathematics, Tampere University of Technology, PO Box 553, 33101, Tampere, Finland 
          \and
          Universit\'e C{\^o}te d'Azur, Observatoire de la C{\^o}te d'Azur, CNRS, Laboratoire Lagrange, France
          \and
         SETI Institute, Carl Sagan Center, 189 Bernado Avenue, Mountain View CA 94043, USA 
         \and
         IMCCE, Observatoire de Paris, 77 avenue Denfert-Rochereau, F-75014 Paris Cedex, France
         \and
         Astronomical Observatory Institute, Faculty of Physics, Adam Mickiewicz University, S{\l}oneczna 36, 60-286 Pozna{\'n}, Poland
         \and
         Institute of Physics, University of Szczecin, Wielkopolska 15, 70-453 Szczecin, Poland
         \and
         Space sciences, Technologies and Astrophysics Research Institute, Universit\'e de Li\`ege, All\'ee du 6 Ao\^ut 17, 4000 Li\`ege, Belgium
         \and 
         Open University, School of Physical Sciences, The Open University, MK7 6AA, UK
         \and
         Jet Propulsion Laboratory, California Institute of Technology, 4800 Oak Grove Drive, Pasadena, CA 91109, USA
         \and
         European Space Agency, ESTEC - Scientific Support Office, Keplerlaan 1, Noordwijk 2200 AG, The Netherlands
        \and
        TMT Observatory, 100 W. Walnut Street, Suite 300, Pasadena, CA 91124, USA
        \and
         Secci\'on F\'isica, Departamento de Ciencias, Pontificia Universidad Cat\'olica del Per\'u, Apartado, Lima 1761, Per\'u
         \and
         European Southern Observatory (ESO), Alonso de Cordova 3107, 1900 Casilla Vitacura, Santiago, Chile
}

   \date{Received x-x-2016 / Accepted x-x-2016}
 
  \abstract
   {Asteroid (7)~Iris is an ideal target for disk-resolved imaging owing to its brightness (V$\sim$7--8) and large angular size of 0.33\arcsec~during its apparitions. Iris is believed to belong to the category of large unfragmented asteroids that avoided internal differentiation, implying that its current shape and topography may record the first few 100~Myr of the solar system's collisional evolution.}
   {We  recovered information about the shape and surface topography of Iris from disk-resolved VLT/SPHERE/ZIMPOL images acquired in the frame of our ESO large program.} 
   {We used the All-Data Asteroid Modeling (\adam{}) shape reconstruction algorithm to model the 3D shape of Iris, using optical disk-integrated data and disk-resolved images from SPHERE and earlier AO systems as inputs. We analyzed the SPHERE images and our model to infer the asteroid's global shape and the morphology of its main craters. }
   {We present the 3D shape, volume-equivalent diameter D$_{{\rm eq}}$=214$\pm$5~km, and bulk density $\rho$=2.7$\pm$0.3~g\cdot cm$^{-3}$ of Iris. Its shape appears to be consistent with that of an oblate spheroid with a large equatorial excavation. We identified eight putative surface features 20--40 km in diameter detected at several epochs, which  we interpret as impact craters, and several additional crater candidates. 
   Craters on Iris have  depth-to-diameter ratios that are similar to those of analogous 10 km craters on Vesta.}
    {The bulk density of Iris is consistent with that of its meteoritic analog based on spectroscopic observations, namely LL ordinary chondrites. Considering the absence of a collisional family related to Iris and the number of large craters on its surface, we suggest that its equatorial depression may be the remnant of an ancient (at least 3\,Gyr) impact. Iris's shape further opens the possibility that large planetesimals formed as almost perfect oblate spheroids. Finally, we attribute the difference in crater morphology between Iris and Vesta to their different surface gravities, and the absence of a substantial impact-induced regolith on Iris. }


  \titlerunning{Asteroid (7)~Iris seen by VLT/SPHERE/ZIMPOL}
  \maketitle

\section{Introduction}\label{sec:introduction}

The largest asteroids (typically with D$\geq$200km) are ideal targets for investigating the collisional history of the asteroid belt. Their outer shell has witnessed 4.6 Gyr of collisions, contrary to that of smaller asteroids, which are fragments of once larger bodies \citep{Morbidelli2009}. The outer shell and overall shape of many of the large bodies, however, has been affected  by external processes (such as impacts) and by internal processes via the radioactive decay of $^{26}$Al. During the first 100~Myr that followed their formation, this early heat source generated enough energy to melt/fluidify their interiors, relax their shapes and surface topography, and erase their primordial cratering records. 
Like the terrestrial planets, these bodies (including (1)~Ceres and (4)~Vesta)  do not thus offer the possibility to decode the early collisional environment of the solar system ($\leq$50--200 Myr after solar system formation). More specifically, \citet{Fu2014} found that Vesta's early collisional record (40--200 Myr after CAIs) was erased during its early relaxation phase. The cratering record on Ceres is even less informative as its surface has been continuously viscously relaxed over the age of the solar system and has also experienced widespread resurfacing, which explains the lack of large craters observed across its surface \citep{Hiesinger2016, Marchi2016}.

This, however, may not be the case for the parent bodies of ordinary chondrites (OCs), namely S-type asteroids. Both thermal evolution models of OC parent bodies  \citep[e.g.,][and references therein]{Monnereau2013} and spectroscopic observations of large S-type families suggest that metamorphosed type 4--6 OCs reflect the internal compositional structure of the largest S-type asteroids \citep{Vernazza2014}. This implies that the interiors of these bodies never melted and impacts have been the only evolution process acting since their formation $\sim$2~Myr after the formation of the solar system \citep[e.g.,][]{Monnereau2013}. Thus, the shapes of the largest S-type objects ($D>$150 km), and to a lesser extent their topography, may help constrain whether the collisional activity was more important during the first 50--200 Myr of the solar system evolution compared to the subsequent $\sim$4.4 Gyr.

As part of our ESO large program (ID 199.C-0074; \citealt{Vernazza2018}), we observed asteroid (7)~Iris (hereafter  Iris) with the VLT/SPHERE/ZIMPOL instrument over a full rotation. Iris, which is one of the four $D$>200 ~km S-type main belt asteroids along with (3)~Juno, (15)~Eunomia, and (29)~Amphitrite
\citep[e.g.,][]{Viikinkoski2015, Viikinkoski2017, Hanus2017b}, is an exceptional target for direct-resolved imaging with adaptive optics (AO) due to its large angular size as seen from the Earth (0.33\arcsec) during opposition. Iris is located in the inner part of the asteroid belt, close to Vesta ($a=2.39$~au, $e=0.23$, $i=5.5\degr$), and possesses an LL-like surface composition \citep{Vernazza2014}. Unlike many of the largest asteroids, Iris is not associated with a dynamical family. At first sight, this seems to imply that it did not suffer a major impact during its recent history \citep[most families identified to date are younger than 2\,Gyr; see][]{Spoto2015}. 

The article is organized as follows. In Sect.~\ref{sec:ao} we describe our observations, the data reduction, and deconvolution. We present the 3D shape model of Iris and its bulk density   in Sect.~\ref{sec:ADAM}. We  analyze of the global shape and surface topography, including a list of the craters,  in Sect.~\ref{sec:shape} and Sect.~\ref{sec:craters}, respectively.  Our 3D shape model is then compared to that of \citet{Ostro2010} derived from and independent dataset of radar data. Finally, we summarize the implications of our work in Sect.~\ref{sec:conclusions}.

\section{Observations}\label{sec:data}

\subsection{Disk-resolved images}\label{sec:ao}

\begin{figure}
\begin{center}
\resizebox{1.0\hsize}{!}{\includegraphics{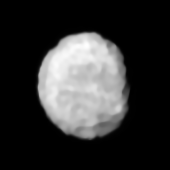}\includegraphics{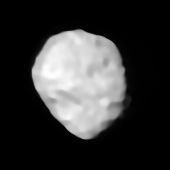}}

\resizebox{1.0\hsize}{!}{\includegraphics{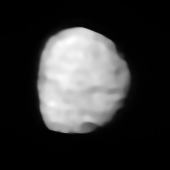}\includegraphics{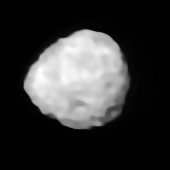}}
\end{center}
\caption{\label{fig:Deconv}VLT/SPHERE/ZIMPOL images of (7) Iris obtained on October 10 and 11, 2017, and deconvolved with the \mistral~algorithm. See Table~\ref{tab:ao} for details on the observing conditions.}
\end{figure}

Five series of images of Iris were acquired with the SPHERE instrument (ESO/VLT; \citealt{Beuzit2008}) during two consecutive nights in October 2017 (see Fig.~\ref{fig:Deconv} and Table~\ref{tab:ao}). The second series from the first night was obtained immediately after the first one, implying that these two epochs sample almost the same geometry (the 7-minute gap between the observations corresponds only to a 5\degr~difference in rotation phase). In order not to overweight this geometry in the shape modeling, we lowered the statistical weight of the first series of images. 
The three remaining series from the second night of observation are of outstanding quality. We clearly resolved several surface features (that we interpret as impact craters) that were consistently present in all images. 
The spectacular quality of the images is driven by the large angular size of Iris of 0.33\arcsec. 
Considering the distance of Iris at the time of our observations, one pixel represents $\sim$2.3 km at its surface. 
From our benchmark study of (4)~Vesta \citep{Fetick2019}, which orbits approximately at the same distance from the Earth, we know that we can reliably identify surface features down to $\sim$8--10 pixels in size. The achieved spatial resolution is  therefore  at least of $\sim$30 mas, corresponding to a projected distance of $\sim$20~km.
We also note that the apparent geometry of Iris during our observations was limited to the southern hemisphere as we observed it almost pole-on (aspect angle $\sim$160\degr). Therefore,  even though the aim of the large
program was to obtain six epochs sampling the full rotation phase,  additional images  would not have brought enough new information to justify acquiring them. 

All VLT/SPHERE observations were obtained by the ZIMPOL instrument \citep{Thalmann2008} in the  narrowband imaging configuration (N$\_$R filter; filter central wavelength = 645.9 nm, width = 56.7 nm). The observing strategy is the same for all targets within our ESO large program \citep[see][for more details]{Vernazza2018}. We used Iris as a natural guide star for AO correction during each series of five cubes of images with a total exposure of 60\,s. All images, with the exception of those from the first series, were collected under the required seeing conditions ($\leq$0.8\arcsec) and an airmass below 1.7. We also observed a nearby star aimed to be used as an estimate of the instrumental point spread function (PSF) for deconvolution purposes. However, we later used a parametric PSF for deconvolution of the asteroid data rather than the observed one, owing to better performances \citep[see also][]{Viikinkoski2018, Fetick2019}. Specifically, we utilized the \mistral~deconvolution algorithm \citep{Fusco2003, Mugnier2004}, and a parametric PSF with a Moffat profile \citep{Moffat1969}. The reliability of using a parametric PSF for deconvolving asteroid images was demonstrated by applying this method to images of (4) Vesta acquired by our program \citep{Fetick2019}: A comparison of these images to {in situ} data collected by the NASA Dawn mission reveals a very good agreement of the surface features detected on Vesta. The deconvolved images of Iris are shown in Fig.~\ref{fig:Deconv}.

In addition to our VLT/SPHERE data, we compiled 19 Keck/NIRC2 and 3 VLT/NaCo images of Iris (see \citealt{Viikinkoski2017}, and  Fig.~\ref{fig:comparison2} and Table~\ref{tab:ao} for additional information). Although the difference in spatial resolution between NIRC2 (pixel scale 9.9 mas), NaCo (12.25 mas), and our VLT/SPHERE images (pixel scale 3.6 mas) is rather large, a subset of these older data still contains valuable information about the shape of Iris. Moreover, they sample additional geometries (aspect angles of $\sim$70\degr) compared to the VLT/SPHERE images, obtained close to a southern pole-on configuration (aspect angle of $\sim$160\degr). Therefore, the NIRC2 and NaCo data essentially provide constraints on the parts of the shape that were not seen by SPHERE, and on the dimension along the rotation axis. Unfortunately, some data were affected by severe deconvolution artifacts, which prevented them from being  used for the shape modeling. We list these images for completeness.

\subsection{Optical disk-integrated photometry}\label{sec:photometry}

Optical lightcurves are particularly important for the spin period determination and a proper phasing of the AO images. We downloaded 39 lightcurves from the online Database of Asteroid Models from Inversion Techniques \citep[DAMIT\footnote{\url{http://astro.troja.mff.cuni.cz/projects/asteroids3D}},][]{Durech2010}, which also contains the most recent shape models of the asteroid Iris. In addition, we utilized 94 single lightcurves covering apparitions in 2006, 2008, 2011, and 2012 extracted from the SuperWASP image archive \citep{Grice2017}. The whole optical dataset samples 17 different apparitions between years 1950 and 2013. The characteristics of the photometric data are listed in Table~\ref{tab:lcs}.

\section{Results}\label{sec:results}

\subsection{3D shape reconstruction, size, and bulk density}\label{sec:ADAM}

\setkeys{Gin}{draft=false}
\begin{figure*}
\begin{center}
\resizebox{0.90\hsize}{!}{\includegraphics{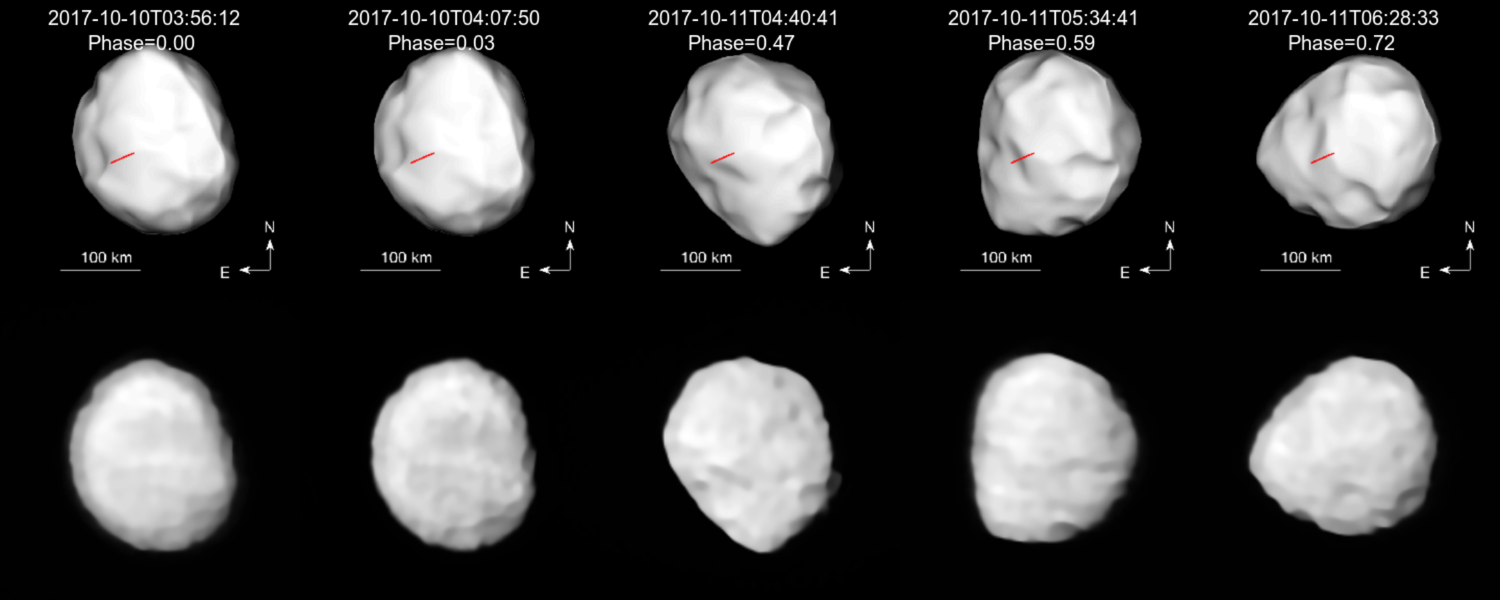}}
\end{center}
\caption{\label{fig:comparison}Comparison between the VLT/SPHERE/ZIMPOL deconvolved images of Iris (bottom panel) and the corresponding projections of our \adam{} shape model (top panel). The red line indicates the position of the rotation axis. We use a nonrealistic  illumination to highlight the local topography of the model. The selected illumination significantly enhances the ridge that spreads from the top to the bottom of the projections on the left-hand side of the figure.} This ridge is a minor modeling artifact caused by the limited number of observing geometries sampled by SPHERE.
\end{figure*}
\setkeys{Gin}{draft=true}

\begin{table*}
 \centering
 \begin{threeparttable}
 \begin{tabular}{llccc}
  \hline
  Parameter               & Unit & \citet{Ostro2010} & \citet{Viikinkoski2017} & This work \\
  \hline
   $D$       & km   &208$\pm$35  & 216$\pm$7     & 214$\pm$5  \\
   $\lambda$ & deg. &15$\pm$5    & 18$\pm$4      & 19$\pm$3  \\
   $\beta$   & deg. & +25$\pm$15   & +19$\pm$4      & +26$\pm$3   \\
   $P$       & h    & 7.1388(1)   & 7.138843(1)   & 7.138843(1) \\
   $a$       & km   & 253$\pm$38  & 262$\pm$10    & 268$\pm$5 \\
   $b$       & km   & 228$\pm$34  & 236$\pm$6     & 234$\pm$4  \\
   $c$       & km   & 193$\pm$39  & 182$\pm$6     & 180$\pm$6  \\
   $a/b$     &      & 1.1$\pm$0.2 & 1.11$\pm$0.05 & 1.15$\pm$0.03  \\
   $b/c$     &      &1.2$\pm$0.3 & 1.30$\pm$0.05 & 1.30$\pm$0.05 \\
   $m$       & $\times 10^{18}$ kg     &12.5$\pm$0.2$^a$ & 12.9$\pm$2.1$^b$  & 13.75$\pm$1.30  \\
   $\rho$    & (g\cdot cm$^{-3}$) &2.66$_{-0.29}^{+0.85}$ & 2.4$\pm$0.5   &  2.7$\pm$0.3 \\
  \hline
 \end{tabular}
 \begin{tablenotes}[para,flushleft]
     \centering $^a$ \citet{2005-SoSyR-39-Pitjeva}, $^b$ \citet{Carry2012b}. 
 \end{tablenotes}
 \end{threeparttable}
 \caption{\label{tab:param}
 Volume-equivalent diameter ($D$), dimensions along the major axis ($a$, $b$, $c$), sidereal rotation period ($P$), spin-axis ecliptic J2000 coordinates (longitude $\lambda$ and latitude $\beta$), mass ($m$), and bulk density ($\rho$) of Iris as determined here, compared with values from the works of \citet{Ostro2010} and \citet{Viikinkoski2017}. Uncertainties correspond to 1\,$\sigma$ values.}
\end{table*}

We used the All-Data Asteroid Modeling (\adam){}  inversion technique \citep{Viikinkoski2015, Viikinkoski2016} for the reconstruction of the 3D shape model and the spin of Iris using disk-integrated (optical lightcurves) and disk-resolved data as inputs. The \adam{} technique is a well-described inversion algorithm that has already been applied to tens of asteroids \citep[e.g.,][]{Viikinkoski2015b, Viikinkoski2017, Viikinkoski2018, Hanus2017a, Hanus2017b, Marsset2017, Vernazza2018}. Exhaustive information about this modeling technique can be found in these studies.

Optical lightcurves are often required for \adam{}, because they stabilize the shape optimization and constrain the parts of the shape not covered by the AO observations. Disk-resolved data provide necessary constraints on the local topography. Without these data, the use of \adam{} would be redundant: standard lightcurve inversion codes would be sufficient. We note that the a priori knowledge of the sidereal rotation period and the spin axis orientation of the asteroid is used as an initial input for \adam{}. In the case of Iris, both quantities were already constrained by previous studies \citep{Kaasalainen2002b, Ostro2010, Viikinkoski2017}. 

We applied the \adam{} algorithm to our dataset of 133 optical lightcurves, 25 VLT/SPHERE/ZIMPOL images from five different epochs, 19 Keck/NIRC2 images, and three VLT/NaCo images. We first computed a rough global representation of the shape model of Iris by enhancing the weight of the lightcurve data with respect to that of the AO images. We ensured that the shape model solution was stable by testing different combinations of (i)~shape support (i.e., octantoids and subdivision) (ii)~AO data types (deconvolved or nondeconvolved images), and (iii)~shape model resolutions. Moreover, we  lowered the weight of the first-epoch AO SPHERE images, as well as that of several Keck images of poor quality. The low-resolution model was then used as a starting point for the modeling with more topographic details. The resulting model contains the most prominent surface features visible in the SPHERE images (Fig.~\ref{fig:comparison}).

Table~\ref{tab:param} provides the final values for the spin-axis orientation, sidereal rotation period, volume-equivalent diameter, and dimensions along the major axes of Iris. These parameters were computed as their average values from the various shape models, and the reported uncertainties correspond to their range of values within these models. Only one shape model was selected as the representative solution that will be included in the DAMIT database. This solution, which is based on the deconvolved AO data and the octantoid shape support \citep{Viikinkoski2015}, has parameter values that are slightly different from the average ones reported in Table~\ref{tab:param}, while within the quoted uncertainties. A comparison between the SPHERE AO images and the corresponding projections of the model is shown in Fig.~\ref{fig:comparison}. A comparison between the Keck/NIRC2 and VLT/NaCo images and the model is provided in Fig.~\ref{fig:comparison2}. 

The shape model contains only the most obvious craters, for which we measured the size and depth (Table~\ref{tab:craters_list}). In general, the crater sizes derived from the model are overestimated (except for Xanthos; see Sect.~\ref{sec:craters}), whereas their depth is underestimated compared to the estimates retrieved directly from the images (Sect.~\ref{sec:craters}). This is an outcome of the modeling technique.

We combined the volume of Iris derived from our shape model (Table~\ref{tab:param}) with its mass estimate derived from all reliable estimates found in the literature \citep[see Table~\ref{tab:mass},][]{Carry2012b}. Specifically, we used the median of the 30 reported mass estimates, after removing the five least reliable values. The resulting mass values are similar to the median of the whole sample. Considering the high number of reliable mass estimates available and their disparity, we consider our mass estimate for Iris to be rather robust: $(13.75 \pm 1.30) \times 10^{18}$ kg (1$\sigma$ error). This leads to a bulk density of $\rho = 2.7 \pm 0.3$~g/cm$^{3}$ (1$\sigma$ error), with an uncertainty that is dominated by the uncertainty on the mass. 

In order to compare the bulk density of Iris with that of other S-type asteroids, we compiled a database of reliable bulk density measurements for 21 S-type asteroids listed in Table~\ref{tab:literature}, and shown in Fig.~\ref{fig:literature}. Iris's bulk density appears slightly lower than that of  most of the other large ($D>150$~km) S-type asteroids, but it is still consistent within the reported error bars. Finally, the density appears consistent with that of its meteoritic analog, namely LL ordinary chondrites (the mean bulk density of LL chondrites is $3.22 \pm 0.22$~g/cm$^{3}$; \citealt{Consolmagno2008}).

\subsection{Global shape}\label{sec:shape}

\setkeys{Gin}{draft=false}
\begin{figure}
\begin{center}
\resizebox{0.493\hsize}{!}{\includegraphics{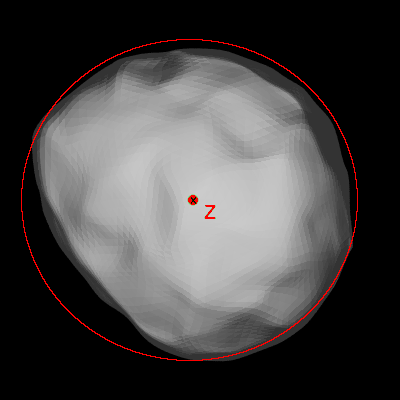}}\resizebox{0.5\hsize}{!}{\includegraphics{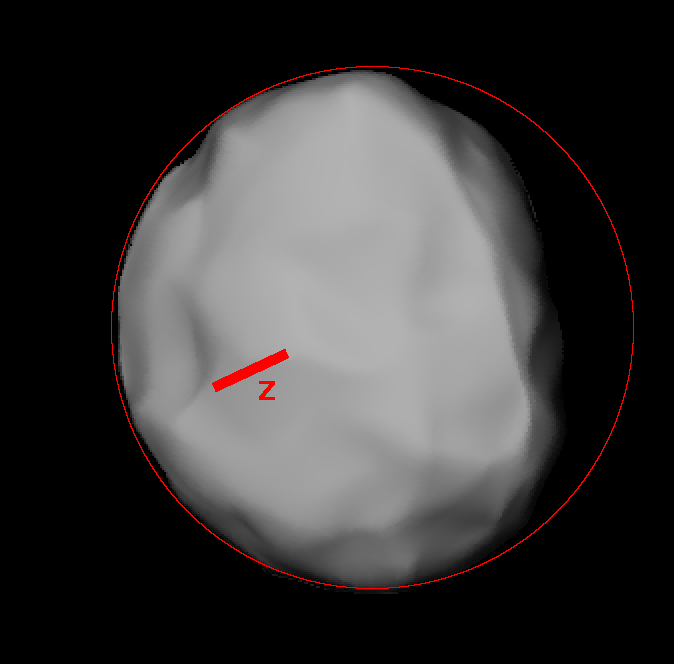}}
\end{center}
\caption{\label{fig:zview}Visualization of the shape model of Iris from a southern pole-on view (top), and from the geometry of the second epoch of observation (rotation phase 0.03, bottom). We embedded the shape projection within a circle to highlight the excavated part and included the z-axis orientation.}
\end{figure}
\setkeys{Gin}{draft=true}


The overall shape of Iris is probably one of the most intriguing properties of this body revealed by our observations. 
Both the SPHERE images and the global appearance of our 3D shape model indicate an oblate spheroidal shape, with a seemingly large excavation close to its equator (Fig.~\ref{fig:zview}). 
Based on our 3D shape model, we estimate that this excavation represents between 10\%\ and 15\% of the volume of Iris.

The nearly spheroidal shape of Iris opens the possibility that this asteroid formed as an almost perfect oblate spheroid. If this is the case, its equatorial depression may be the result of a large-scale impact. Yet, this seems surprising considering the absence of a collisional family associated with Iris (Appendix~\ref{sec:family}). 
Families associated with large asteroids are a common feature in the main belt, and some of these families are known to exist over several Gyr before they dissipate via Yarkovsky drifting and collisional grinding \citep{Vokrouhlicky2006d,VokrouhlickyAIV2015}.
For instance, the NASA Dawn mission revealed ages of 1 and 2~Gyr for the two large craters that form the Rheasilvia basin on Vesta \citep{Schenk2012, Marchi2012}, implying similar ages for the two collisional families associated with Vesta \citep{Milani2014}. 
Additional old families in the asteroid belt include those of
Hygiea (2~Gyr), Eunomia (2.5~Gyr), Koronis (2.5~Gyr), and Themis (3~Gyr)
\citep{Broz2013b, Nesvorny2015,Spoto2015}. 
Assuming that a large impact is at the origin of the depression, the lack of an Iris family seems to imply that the collision occurred during the very early phase of solar system evolution ($>$3~Gyr ago). 
This is strengthened by the fact that Iris is located in a dynamically stable region of the asteroid belt, nearby Vesta, and far away from any strong orbital resonances with the giant planets. This location implies that its family must have depleted very slowly. Numerical integrations for the collisional evolution of the asteroid belt reveal that events able to excavate at least 10\% of the mass of Iris happen 0 to 2 times over a timescale of 4~Gyr (see Fig.~\ref{fig:MB_Iris_ONLY_hist_1PERCENT} and Appendix~\ref{sec:mira}).

An alternative explanation for the lack of an Iris family could be that Iris experienced, even recently, a near-miss ``hit-and-run'' collision where it was impacted close to the edge. In this case, the fragments would have gained most of the projectile momentum and would have scattered away from the space of the Iris proper orbital elements.


Finally, we cannot entirely rule out that Iris may have never had a spheroidal shape, and that its current shape is close to the original one. Our program will help determine whether similar shapes are found for other D$\sim$200\,km asteroids. In turn, this will provide new constraints on the origin of asteroid shapes, and the possible link existing between asteroid shapes and the presence of collisional families.

\subsection{Impact craters}\label{sec:craters}

\begin{table*}
 \centering
 \begin{tabular}{lccccccccccc}
  \hline
  Crater      & $\lambda$ & $\phi$ &  \multicolumn{4}{c} {Rotation phase} & Diameter~(km) & Diameter~(km) & Depth~(km) & Depth~(km) & $d/D$ \\
              &    (\degr)&(\degr) & 0.03 & 0.47 & 0.59 & 0.72 &(image) & (model) & (image) & (model) & (image) \\
  \hline
  \hline
   Xanthos    & 0  & -16 & Y    & Y    & Y    & Y    & 38$\pm$5  & 44$\pm$5  & 13$\pm$3  & 12$\pm$2   & 0.34$\pm$0.09 \\
   Erythros   & 255& -22 & -    & Y    & Y    & Y    & 31$\pm$8  & --        & --  &   --       & -- \\
   Cyanos     & 53 & -25 & -    & Y    & Y    & Y    & 25$\pm$5  & --        & --        &  --        & -- \\
   Chloros    & 47 & -40 & n.d. & Y    & Y    & Y    & 41$\pm$6  & 62$\pm$5  & --        & 6$\pm$2    & -- \\
   Cirrhos    &  2 & -34 & n.d. & Y    & Y    & Y    & 23$\pm$5  & 31$\pm$5  & --        &  2$\pm$2   & -- \\
   Porphyra   & 202& -21 & Y    & Y    & Y    & Y    & 35$\pm$5  & --        & --        & --         & -- \\
   Chrysos    & 182& -14 & Y    & n.i. & n.i. & Y    & 36$\pm$9  & 37$\pm$5  & --  & 3$\pm$2    & -- \\
   Glaucos    & 164& -20 & Y    & n.i. & n.i. & Y    & 30$\pm$9  & 49$\pm$5  & --        & 6$\pm$2   & --  \\
   A          &  65& -21 & n.d  & A    & n.d. & n.d. & 23$\pm$5  & --        & 5$\pm$3   & --         & 0.22$\pm$0.14 \\
   B          & 126& -2 & Y     & -    & -    & n.d. & 35$\pm$5  & --        & 6$\pm$3   & --         & 0.17$\pm$0.09 \\
   C          & 67 & -65 & n.i. & Y    & n.i. & n.i. & ~(64$\times$16)$\pm$5 & -- & --   & --         & -- \\
   D          & 105& -27 & n.i. & Y    & n.d. & n.i. & 21$\pm$5  &  --       & --        & --         & -- \\
   E          &  57& -15 & Y    & -    & -    & -    & 23$\pm$5  &  --       & --        & --         & -- \\
   F          & 18 & -19 & -    & Y    & Y    & n.i. & 27$\pm$5  &  --       & --        & --         & -- \\
   G          & 243& -5  & -    & Y    & Y    & Y    & 43$\pm$8  &  --       & --        & --         & -- \\
\hline 
 \end{tabular}
 \caption{\label{tab:craters_list}Identified and suggested topographic features on Iris. We indicate for each epoch whether we clearly identified the feature (Y);  if we probably identified it, but the contours were not well defined (n.d.);  if we did not identify it (n.i.); or if the feature does not fall within the visible part of the surface (-). The uncertainties in the diameter ($D$) and depth ($d$) values were computed as their variance across the complete set of images in which they are visible, unless the variance is lower than 3\,km. In that case the uncertainty was set to 3\,km, which reflects a conservative approximation of the pixel size at the distance of Iris ($\sim$2.3~km). Moreover, we also included crater diameters and depths measured on the shape model. Their uncertainties reflect the resolution of the shape model.
 }
\end{table*}

\setkeys{Gin}{draft=false}
\begin{figure*}
\begin{center}
\resizebox{\hsize}{!}{\includegraphics{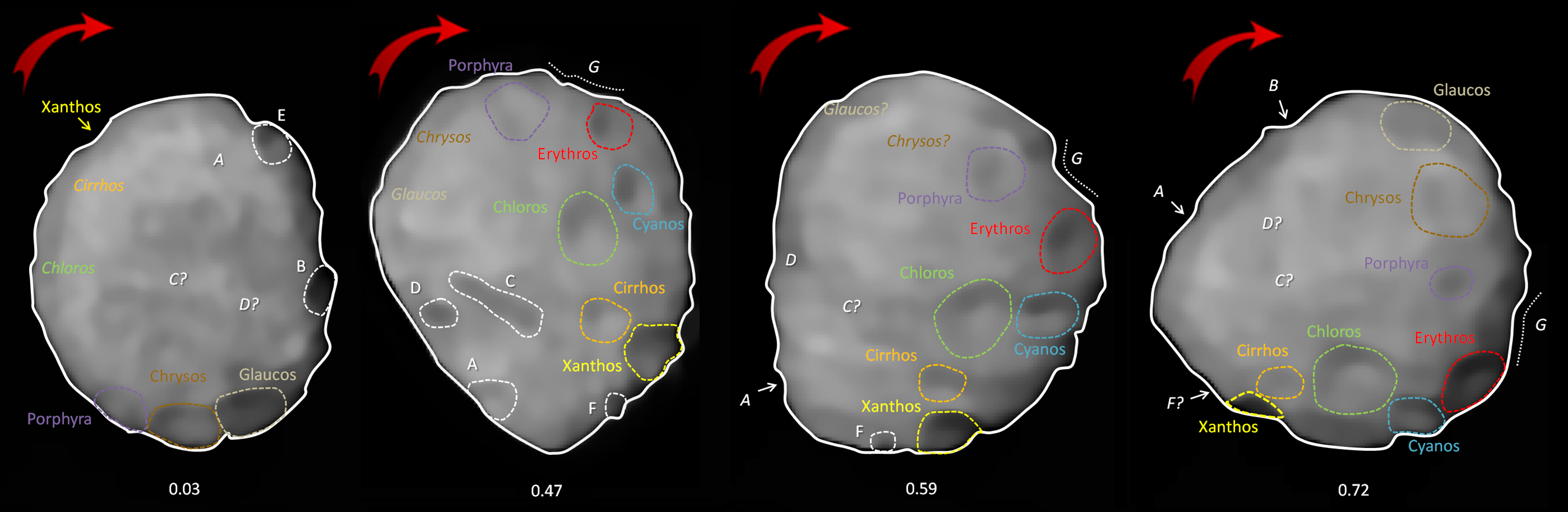}}
\end{center}
\caption{\label{fig:craters}Identified and proposed surface features on Iris. Putative impact craters are indicated by names and often with a contour showing the estimated area they cover. Candidates for impact
craters are indicated by white-font letters and contours, in italic if the contours were not clearly
defined. A question mark indicates the expected position of a feature that could not be identified on the image, likely due to unfavorable illumination.}
\end{figure*}
\setkeys{Gin}{draft=true}

We identified eight topographic features on the images, with typical diameters between 20--50\,km, that we consider as putative impact craters based on their apparent circular morphology (Table~\ref{tab:craters_list}, Fig.~\ref{fig:craters}). We nicknamed them with the Greek names of colors to reflect their association with Iris, the goddess of the rainbow in  Greek mythology. The pole-on geometry of Iris during our observations allowed us to accurately trace the position of the craters throughout a complete rotation period. All the reported craters were identified from the images, simultaneously using the shape model to track their location at every epoch in order to verify their reliability and visibility. 

Six out of the eight identified putative craters are clearly visible in at least three epochs shown in Fig.~\ref{fig:craters}. The remaining two craters were identified at a single epoch that corresponds to the  Iris 0.03-rotation phase angle. Of these two craters, the one we call Chrysos is also visible at the 0.72-rotation phase angle. These two craters remained undetected in the other images, likely due to unfavorable illumination conditions \citep[see also][]{Fetick2019}. We indicate their expected location on each image based on the rotation period of Iris, and using their position on the single image where they were detected as a benchmark. Several additional surface features considered as potential candidates for impact craters are  highlighted in Fig.~\ref{fig:craters} by the letters A--G.

Table~\ref{tab:craters_list} summarizes the coordinates, estimated size, and proposed names or designations for the identified topographic features. The reference zero longitude of the asteroid-centric coordinate system was defined as the location of the large equatorial depression Xanthos, following the IAU recommendation for the reference frame \citep{Archinal2018}. For each epoch we also
indicate  whether  the feature falls within the visible part of the asteroid and whether  it could be identified on the series of images that corresponds to that epoch. Crater diameters were directly measured on the images by first removing the illumination gradient from the asteroid images \citep[see][for details]{Carry2008, Carry2010a}, and then by drawing a projection of the image's brightness level (number of counts) along the craters and for different orientations. The edges of the craters were defined as the locations where the profiles start to reach a plateau outward from the center of the crater. We used the 3D shape model to measure the planetocentric latitude and longitude of the craters. For the five craters that are visible on the shape model, we computed their diameter and depth following the method described in \citet{Vernazza2018}. The values derived for Xanthos and Chrysos are consistent with the measurements from the images, while those for Chloros, Cirrhos, and Glaucos are overestimated by about 50\%. Moreover, we measured the depth of three craters conveniently located near the asteroid's terminator on the images acquired at 0.72-rotation phase angle. This configuration allows us to measure the orthogonal distance from the bottom of the crater to the tangent of the surface between the rims. Specifically, we find a value of $\sim$15~km for Xanthos, with an uncertainty of 3~km that reflects the pixel scale of the images. A similar depth for Xanthos was also derived  from the shape model. This value seems to be robust because the Xanthos appearance on the shape model is largely consistent with the images.

In the case of Xanthos, the measured depth-to-diameter ($d/D$) ratio is  $\sim$0.34. This value is higher than that for similarly sized craters on Vesta \citep[$d/D$=0.15--0.27,][]{Vincent2014}. Given that Iris and Vesta are rocky bodies made of similar tensile-strength materials (LL ordinary chondrites and HED achondrites, respectively), the observed difference in crater morphology may be due to Vesta's larger surface gravity $g$ (around 3 times higher than that of Iris), which would more efficiently reshape its craters, for example through landslides and erosion. Vesta's larger $g$ also makes it more likely to refill its craters with a reaccreted, impact-generated surface regolith, whereas a larger fraction of impact ejecta would be lost after an impact on a smaller, Iris-sized body. Along these lines, \citet{Thomas1999b} found that the crater morphology on asteroids and the satellites of the giant planets scales as the inverse of surface gravity, therefore, 30 km craters on Iris should have (and actually have) similar morphology as 10 km craters on Vesta.

Apart from Vesta, the only rocky asteroid imaged by a spacecraft where a crater  20--60~km in diameter can be found is (21) Lutetia, a $\sim$100 km large  main belt asteroid. Massilia, the largest crater on Lutetia, has a diameter of 55 km and a depth of about 2 km \citep{Cremonese2012}, which implies $d/D=0.04$. This value is significantly lower than the one we derived for Xanthos. Considering the morphology scaling with the inverse of surface gravity, Massilia should be compared to 20--30 km craters on Iris. Clearly, Massilia is rather shallow compared to the craters on Iris, which could be explained by physical resurfacing processes such as regolith deposits and relaxation. The fact that Massilia's diameter is comparable to the size of Lutetia also certainly impacts its morphology.

We also did not see any evidence of complex craters on Iris. In particular, none of the identified putative craters exhibits the presence of a central peak. At first glance, craters with central peaks may be visible on the images acquired at 0.03-rotation phase angle. However, these features all remain undetected on the other images, which are all of slightly higher quality (acquired at lower airmass). We therefore attribute them to instrumental or deconvolution artifacts. Following the scaling law proposed by \citet{Asphaug1996}, the transition diameter from simple to complex craters for Iris is $D=0.8 Y/(g\,\rho)=74$~km, where $Y=2.10^{7}~{\rm N.m^{-2}}$ is the average tensile strength of silicates, $g=0.08~{\rm m.s^{-2}}$ is the average gravitational acceleration on Iris, and $\rho=2.7~{\rm g.cm^{-3}}$ is its density. The lack of complex craters and central peaks is therefore not surprising considering the range of crater sizes detected on Iris. 

Finally, the observed crater density on Iris is similar to that predicted by numerical simulations (see Fig.~\ref{fig:MB_Iris_ONLY_hist_1PERCENT} and Appendix~\ref{sec:mira}) and consistent with the surface age $>$3 Gyr (i.e., there was not a recent collision that erased the cratering record on Iris). This highlights that models simulating the collisional evolution of the asteroid belt have become robust.

\subsection{Comparison with (4)~Vesta and (21)~Lutetia}

In order to compare the cratering record on the surface of Iris with records on (4)~Vesta and (21)~Lutetia, we computed the corresponding crater density $n$ on Iris. For the six craters with $D_{\rm c} > 30\,{\rm km}$ observed on Iris and the surface area corresponding to a~$220\,{\rm km}$ sphere, i.e., slightly larger than for an equivalent volume, we obtained $n = 3.9\cdot10^{-5}\,{\rm km}^{-2}$. This value is clearly a lower limit because (i)~only about a half of the surface was observed by SPHERE, (ii)~there might still be some observational bias, and (iii)~the cratering record could be affected by  resurfacing. This would increase $n$ by corresponding factors: $f_{\rm visible}$, $f_{\rm bias}$, and $f_{\rm resurf}$. Naturally, the age of the surface is counted from the last catastrophic or reaccumulation event.

For comparison, (4) Vesta and its Rheasylvia (RS) basin floor has a crater density $n_{\rm RS} = 1.7\cdot10^{-5}\,{\rm km}^{-2}$ and an estimated age of $t_{\rm RS} \doteq 1\,{\rm Gyr}$ \citep{Marchi2012,Marchi2015}. If we assume the factors
$f_{\rm visible} = 2$,
$f_{\rm bias} = 1$, and
$f_{\rm resurf} = 1$
for simplicity, the age of the surface of Iris would theoretically be $t \doteq f_{\rm visible} f_{\rm bias} f_{\rm resurf}\, (n/n_{\rm RS})\, t_{\rm RS} \gtrsim 4.0\,{\rm Gyr}$.
More precisely, we should use the $\pi$-scaling factor of \citet{Melosh1989} for the projectile-to-crater ($D_{\rm c}$-to-$d_{\rm p}$) size scaling (see Appendix~\ref{sec:mira} and Eq.~\ref{eq:piscaling}): for the same projectile population with $d_{\rm p} \ge 2\,{\rm km}$, we should have crater diameters $D_{\rm c} \ge 19.3\,{\rm km}$ on Vesta (we neglect the minor differences in collisional probabilities, and gravitational focusing factors). Finally, we also considered the heavily cratered terrains (HCT) on Vesta, for which the crater density is $n_{\rm HCT} \doteq 9.5\cdot10^{-5}\,{\rm km}^{-2}$. Assuming this unit is as old as $4.0\,{\rm Gyr}$, it would lead to $t = 3.3\,{\rm Gyr}$ for the age of the surface of Iris.

On the other hand, (21) Lutetia and its Achaia region has a crater density of $n_{\rm Ach} \doteq 2\cdot10^{-4}\,{\rm km}^{-2}$. This value is substantially higher than for Iris and Vesta; however, it is extrapolated from $D_{\rm c} = 20\,{\rm km}$
with an uncertainty of at least a factor of~2 (there are actually no 30 km craters on Achaia). The corresponding surface age is up to $3.8\,{\rm Gyr}$, i.e., presumably formed during the late heavy bombardment \citep{Marchi2012b}. The $\pi$-scaling factor is very similar to that for Iris. However, a statistics of small numbers plays an important role in this case. Consequently, the inferred age is very uncertain, it may be lower than 1\,{\rm Gyr}, or reach up to 3\,{\rm Gyr} (1$\sigma$).

\subsection{Comparison with the radar model of \citet{Ostro2010}}

An exceptional dataset of Iris delay-Doppler images (the best dataset for a main belt asteroid) was acquired in November 2006 by the Arecibo observatory when Iris was at a distance of only 0.85 AU from the Earth \citep{Ostro2010}. By coincidence, our SPHERE observations were obtained at an  observing geometry very similar to that of  the radar data, so both datasets map only the southern hemisphere of Iris. This is very convenient for the comparison of the performance of the two independent modeling approaches. A radar-based shape model of Iris was reconstructed by \citet{Ostro2010} and kindly provided in the form of a polyhedron by Chris Magri. Unfortunately, the uncertainty in the rotation period from \citet{Ostro2010} was too large to compute the correct rotation phase of the Iris radar model at the time of the SPHERE observations in 2017. Therefore, we found the optimal rotation offset by visually comparing the radar projections for different values of the rotation phase to the projections of the \adam{} model and the SPHERE images. We identified several common surface features on both models and images, so the estimated rotation phase of the radar shape model seems to be reliable at the level of $\sim$5 degrees. 

There are substantial discrepancies between the radar and the AO images as shown in Fig.~\ref{fig:comparison}. Specifically, the contours of the radar shape model projections do not agree well with those of the images or \adam{} model, which is most apparent in epochs 3--5 (rotation phases 0.47, 0.59, and 0.72). Considering that deconvolution of disk-resolved images obtained with AO-fed cameras have been validated by several spacecraft encounters \citep[e.g.,][]{Witasse2006, Carry2008, Carry2010b, Carry2012, Russell2016}, we conclude that the  appearance of the radar shape model is not accurate. On the other hand, several surface features (craters Xanthos, Chloros, Cyanos, and Erythras) identified in the SPHERE images are present in the radar-based model. The concavity related to the Xanthos crater is very prominent in the radar model and  is significantly enhanced compared to the AO images or the \adam{} model. In general, the topography of the radar-based model is more dramatic and exaggerated compared to the SPHERE images, and it contains some spurious features (e.g., sharp mountains) not detected in our images. However, the fact that the main surface features seen in the SPHERE images are also described by the radar-based model illustrates a partial robustness of the modeling technique. 

Finally, the radar model contains a large concavity in the C region that we identified only in one AO epoch (rotation phase 0.47). This further supports the existence of this proposed candidate for an impact feature. 
The radar model indicates a single impact basin, while the image and the \adam{} shape model tends to be more consistent with two partially overlapping craters.

\section{Summary}\label{sec:conclusions}

We obtained VLT/SPHERE/ZIMPOL images of (7) Iris revealing surface details for this object with unprecedented precision. Our set of images was used to constrain and characterize Iris's 3D shape (hence volume and density when combining the latter with current mass estimates), identify several impact craters, measure their sizes, and (for some) their depth. The nearly pole-on orientation of Iris during our observations allowed us to track impact surface features throughout its rotation and to easily discriminate real features from instrumental and deconvolution artifacts, while highlighting the level of reliability of features identification with SPHERE.

The derived bulk density for Iris ($2.7\pm$0.3~g\cdot cm$^{-3}$) appears consistent with its LL ordinary chondrite surface composition. The shape of Iris is reminiscent of an almost perfect oblate spheroid with a large equatorial excavation. This may suggest that Iris formed with a spheroidal shape and subsequently suffered a large impact. In that case, the lack of an asteroid family associated with Iris would imply that this impact occurred a long time ago ($>$3~Gyr). Alternatively, this excavation may be the result of a more recent near-miss hit-and-run collision, in which the fragments were scattered far away from the space of the proper orbital elements of Iris. However, the surface age of $\sim$3--4 Gyr inferred from the crater density makes the recent collision unlikely. It should be noted that we cannot entirely rule out the possibility that this apparent excavation reflects the original shape of Iris. Additional AO observations of large asteroids with and without families will help tackle this question.

Our shape model appears to be only partially consistent with the radar-based shape model of \citet{Ostro2010}: Both models contain similar surface features that we associated with the largest impact craters on Iris covering its southern hemisphere and equatorial region. However, the contours of the radar shape model projections do not agree well with those of the images and the \adam{} model. The partial agreement between the two shape-reconstruction methods based on two independent datasets supports the reliability of the surface features interpreted here as putative impact basins, and suggests that the radar-based shape model tends to overestimate the surface topography and to  reproduce less accurately the global appearance of Iris.

Finally, we attribute the difference in the morphology ($d/D$) of similarly sized craters ($d\sim$30--50 km) between Iris and Vesta (both rocky bodies are made of similar tensile-strength materials) to their different surface gravity, and the absence of a substantial impact-induced regolith on Iris.

\begin{acknowledgements}
We thank Simone Marchi for his pertinent and constructive remarks. JH and JD were supported by  grant 18-09470S from the Czech Science Foundation. This work has been supported by the Czech Science Foundation through grants 18-09470S (J. Hanu\v s, J. \v Durech) and 18-04514J (M. Bro\v z) and by the Charles University Research program No. UNCE/SCI/023. P.~Vernazza, A.~Drouard, J.~Grice, and B.~Carry were supported by CNRS/INSU/PNP. We thank Chris Magri for kindly providing the radar shape model published in \citet{Ostro2010}.
\end{acknowledgements}

\bibliography{benoit,mybib}
\bibliographystyle{aa}

\begin{appendix}

\section{Lack of an Iris family}\label{sec:family}

\setkeys{Gin}{draft=false}
\begin{figure}
\centering
\resizebox{\hsize}{!}{\includegraphics{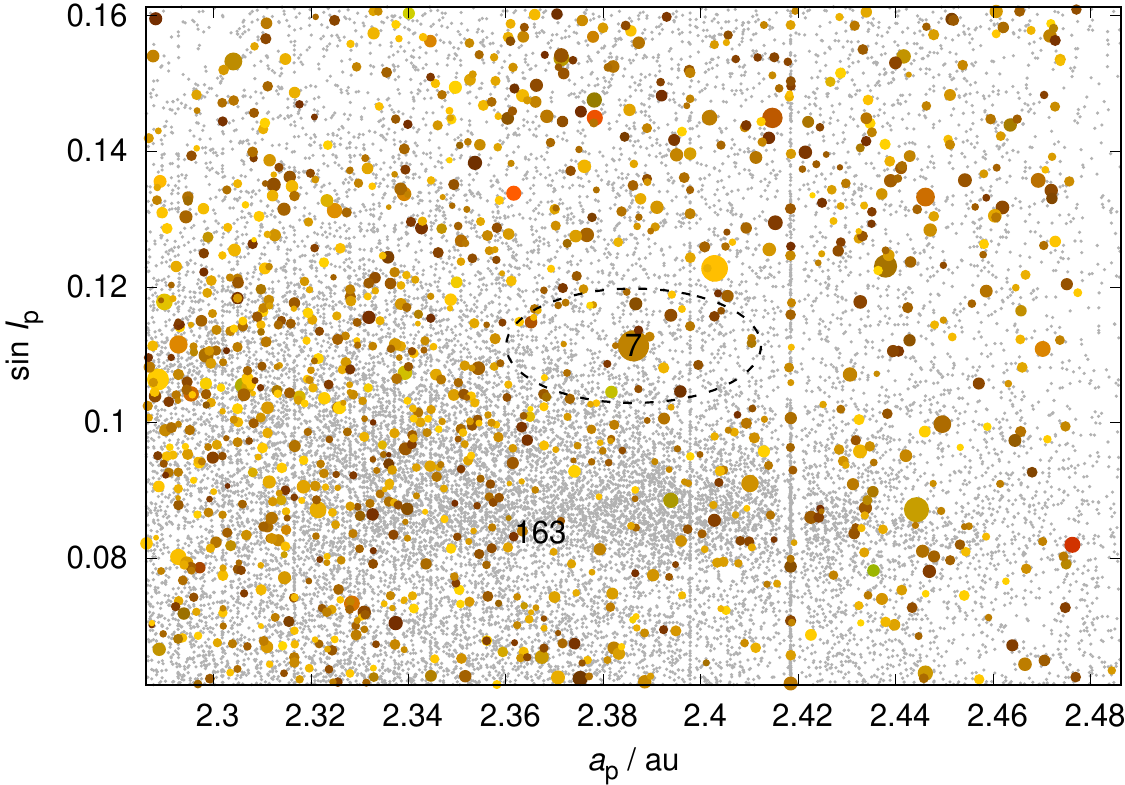}}
\caption{\label{fig:ai_wise_PV0.252}Vicinity of (7)~Iris in the $a_{\rm p}$ vs. $\sin I_{\rm p}$ space of proper orbital elements. The eccentricity range is $e_{\rm p} \in (0.162; 0.262)$. Asteroids with albedo within $\pm0.1$ of the Iris value
($p_V = 0.252$), are plotted in yellow, with symbol size proportional to the body diameter. Asteroids with unknown albedo are plotted as gray dots. The family visible at $\sin I_{\rm p} \doteq 0.09$ is related to the dark C-type asteroid (163)~Erigone. The dashed ellipse corresponds to the escape velocity from Iris, $v_{\rm esc} \doteq 130\,{\rm m}\,{\rm s}^{-1}$, converted to orbital elements using the Gauss equations. This velocity was computed for true anomaly $f = 180^\circ$ and argument of perihelion $\omega = 0^\circ$.}
\end{figure}
\setkeys{Gin}{draft=true}

We demonstrate here the absence of an apparent collisional family linked to Iris. Iris is located in the inner main belt
at $a_{\rm p} = 2.386\,{\rm AU}$, $e_{\rm p} = 0.213$, and $\sin I_{\rm p} = 0.111$. 
From the latest  (September 2018) catalog of proper elements from  \citet{Knezevic2003}, we consider the subset of asteroids with albedo similar to that of Iris, $p_V = 0.252$ (to within $\pm 0.1$), and  confirm that there is no clustering of asteroids in the orbital parameter space of Iris (Fig.~\ref{fig:ai_wise_PV0.252}).
In the vicinity of Iris, there is only one identified family located at lower inclinations and related to the C-type asteroid (163)~Erigone. From a dynamical point of view, there are no strong resonances near Iris except for the three-body resonance 4:2:1 with Jupiter and Saturn at $2.4\,{\rm AU}$.  The Jupiter 3:1 mean-motion resonance is farther away from Iris.

\section{Collisional evolution of Iris}\label{sec:mira}

\setkeys{Gin}{draft=false}
\begin{figure}
\centering
\resizebox{1.0\hsize}{!}{\includegraphics{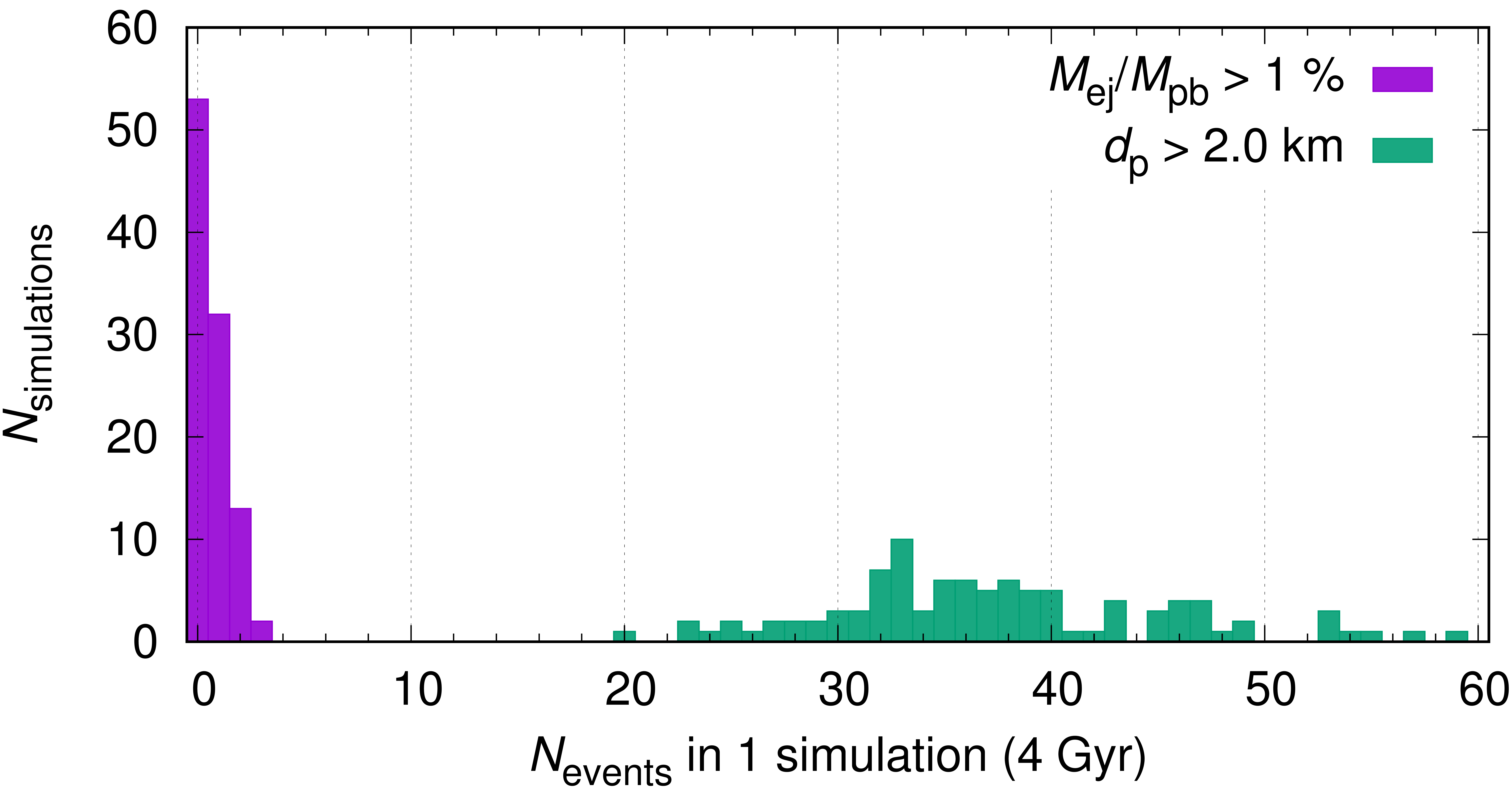}}
\caption{\label{fig:MB_Iris_ONLY_hist_1PERCENT}Number of simulations in which particular events were detected:
(i)~the ejected mass $M_{\rm ej}$ over the parent-body mass $M_{\rm pb}$ was $> 1\,\%$;
(ii)~the projectile diameter was $d_{\rm p} > 2.0\,{\rm km}$.
The former is related to the shape of (7) Iris, the latter to the number of impact craters $>$30 km on Iris.}
\end{figure}
\setkeys{Gin}{draft=true}

The collisional evolution of Iris was modeled using the Boulder code for Monte Carlo collisional simulations
\citep{Morbidelli2009, Cibulkova2014}, in a similar way to that described in \citet{Vernazza2018} for asteroid (89)~Julia. We describe here only the main steps of the procedure, while additional details can be found in the above-mentioned study. 

We considered two populations of objects: the main asteroid belt, and Iris and its collisional fragments.
We assumed a constant intrinsic collisional probability of $P_{\rm i} = 3.10\times10^{-18}\,{\rm km}^{-2}\,{\rm yr}^{-1}$, and impact velocities of $v_{\rm imp} = 5.28\,{\rm km}\,{\rm s}^{-1}$ \citep{Dahlgren1998}.
A size-dependent dynamical decay was included, with rates taken from \citet{Bottke2005}. We used the standard scaling law of \citet{Benz1999} for the kinetic energy threshold $Q^\star_{\rm D}(r)$ of basalt at impact velocities of $5\,{\rm km}\,{\rm s}^{-1}$. The density of Iris and its fragments was set to
$\rho = 2.7\,{\rm g}\,{\rm cm}^{-3}$. 
The nominal time span of the simulations was set to $4\,{\rm Gyr}$. Multiple simulations with different random seeds were performed, allowing fractional probabilities for breakups of the individual large asteroids. Initial conditions were chosen to match the observed size-frequency distribution of the asteroid belt.

Running a hundred simulations, we found that the evolved size-frequency distribution of the asteroid belt closely matches the observed distribution, except for the short-end tail ($D < 1\,{\rm km}$), which  is affected by observation incompleteness. The evolved Iris population contains some fragments, with the largest fragment typically ranging between 5~and 10\,km in size. This is not enough to expect an observable family at any time because small asteroids quickly disperse via the Yarkovsky effect on timescales of a few $100\,{\rm Myr}$. Moreover, we found that about 10\,\% of the simulations barely produced any observable fragments, with the largest fragment having $D < 2\,{\rm km}$. 
Next, we performed an extraction of impact events relevant to the craters seen on Iris. 
We used the $\pi$-scaling factor of \citet{Melosh1989} for the projectile-to-crater ($D_{\rm c}$-to-$d_{\rm p}$) size  scaling,

\begin{equation}\label{eq:piscaling}
D_{\rm c} = C_D\left({1.61gd_{\rm p}\over v_{\rm imp}^2}\right)^{-\beta} \left({m_{\rm p}\over\rho_{\rm t}}\right)^{1\over 3} \sin^{1\over 3}\phi,
\end{equation}
where
$C_D = 1.6$ and $\beta = 0.22$ are the material parameters for competent rock or saturated soil \citep{Schmidt1987}, $g$~is the gravitational acceleration of the target, $m_{\rm p}$~is the projectile mass, $\rho_{\rm p} = 3000\,{\rm kg}\,{\rm m}^{-3}$ its density, $\rho_{\rm t}$ the target density, and $\phi = 45^\circ$ the assumed mean impact angle. Using this scaling law, the measured sizes of the craters seen on Iris ($D_{\rm c} = 20-40\,{\rm km}$) imply projectile sizes of $d_{\rm p} = 1.2$ to $2.9\,{\rm km}$.

The number of impact events with $d_{\rm p} > 2.0\,{\rm km}$ is shown in Fig.~\ref{fig:MB_Iris_ONLY_hist_1PERCENT}. We selected this size of impactor because it corresponds to a crater size of $\sim$30 km. Such craters and larger ones were easily identified in the images, contrary to the smaller ones, which are on the detection limit and their true number should be significantly higher than suggested by our analysis. The median number of events is $\sim$35, with 10\% and 90\% percentile of 25~to 48, which is larger than the observed number of craters of $\sim$10 (including the candidates). However, our detected craters lie only on the southern hemisphere; therefore, their expected number should be larger by a factor of 2. Moreover, we likely did not identify all the craters on the surface, due to unfavorable illumination \citep[see][where we discuss this in more detail]{Fetick2019}. Finally, our simulations neglect all kinds of resurfacing and crater degradation by the ejecta, which should further decrease the number of detected craters. As a result, the number of detected craters is  consistent with the numerical model, which makes the surface age of $>$3 Gyr plausible.

We then focused on impacts that are neither cratering events nor catastrophic disruptions. Such impacts can significantly alter the global shape of the original Iris parent body, and explain the observed depression on Iris (Section~\ref{sec:shape}). Our simulations reveal that events able to produce an ejected mass over the parent body mass ratio of $M_{\rm ej}/M_{\rm pb} > 1\,\%$, which corresponds to the observed excavated volume of the depression on Iris (i.e., $\sim$10\% of the Iris' volume), happen usually once or twice in 4\,Gyr in about half of the simulations (Fig.~\ref{fig:MB_Iris_ONLY_hist_1PERCENT}). Here we assume that the ejected mass of about 10\% of the excavated mass for Iris is greater than for Julia \citep[2\%,][]{Vernazza2018}, however, still reasonable because the proposed impact on Iris is larger than on Julia. Clearly, the ejected mass compared to the excavated mass increases with the impactor size (it is zero for small impactors and equals the excavated mass for catastrophic disruptions).
Therefore, we cannot exclude the possibility that Iris has suffered such a large collision in the past.

While this appears incompatible with the lack of a collisional family associated with Iris, we note that a nearly missed collision, where the target is hit close to its edge, has a similar probability of  happening  as a direct hit: 
for a projectile that delivers at least half of its kinetic energy, the effective cross section is $\pi R_{\rm t}^2$. 
For a projectile that almost misses the target, it is $\pi(R_{\rm t}+r_{\rm p})^2-\pi R_{\rm t}^2$. These areas become equal for $r_{\rm p} = (\sqrt{2}-1)\,R_{\rm t}$. In such a collision, the fragments would gain enough momentum to be scattered far away from the impacted body.

Considering that we observe a nonnegligible number of large asteroids ($D > 100\,{\rm km}$) associated with families \citep{Nesvorny2015}, we expect that some nearly missed objects must also exist in the main belt. Placing Iris in the context of a substantial sample of similarly sized asteroids will help determine whether the near-miss  collision constitutes a viable process to explain the shape of these objects, and the absence of families associated with them. 
This will be possible in the near future when the sample of asteroids with AO-resolved topography  reaches several tens of objects.

\section{Additional figures and tables}

\setkeys{Gin}{draft=false}
\begin{figure*}
\begin{center}
\resizebox{0.99\hsize}{!}{\includegraphics{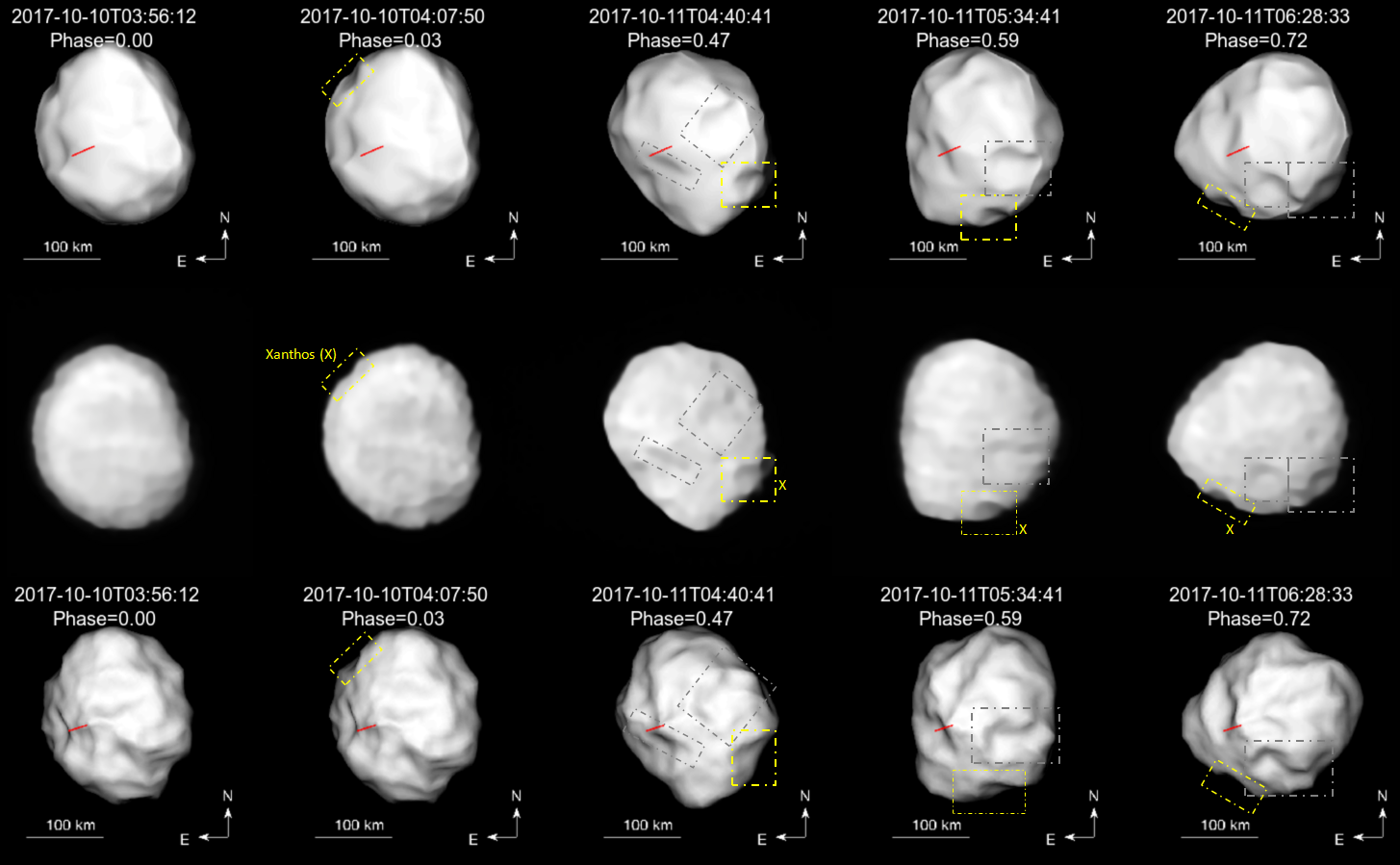}}
\end{center}
\caption{\label{fig:comparison1}Comparison between the VLT/SPHERE/ZIMPOL deconvolved images of Iris (middle panel) and the corresponding projections of our shape model (top panel), and the shape model of \citet{Ostro2010} based on delay-Doppler data collected with Arecibo (bottom panel). The red line indicates the position of the rotation axis. We used a nonrealistic  illumination to highlight the local topography of the models. We highlight the main topographic features.}
\end{figure*}
\setkeys{Gin}{draft=true}

\setkeys{Gin}{draft=false}
\begin{figure*}
\centering
\resizebox{0.9\hsize}{!}{\includegraphics{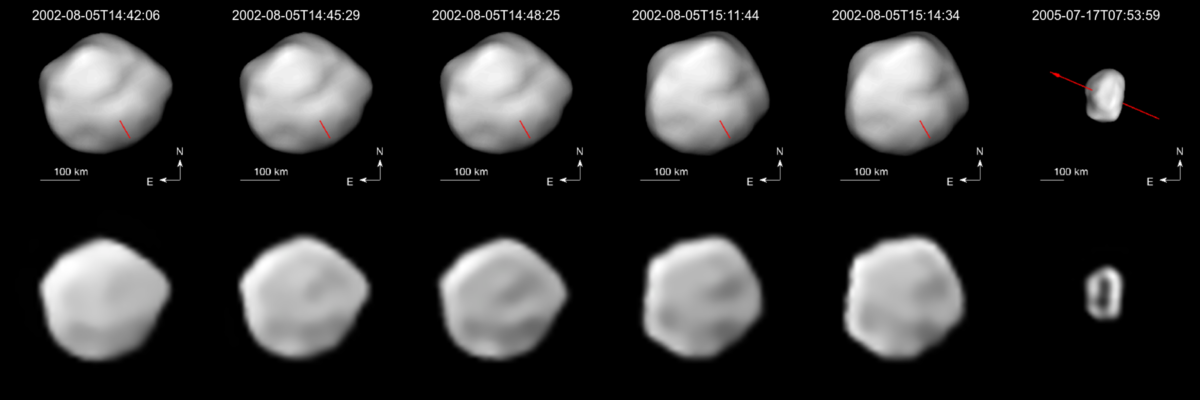}}

\resizebox{0.9\hsize}{!}{\includegraphics{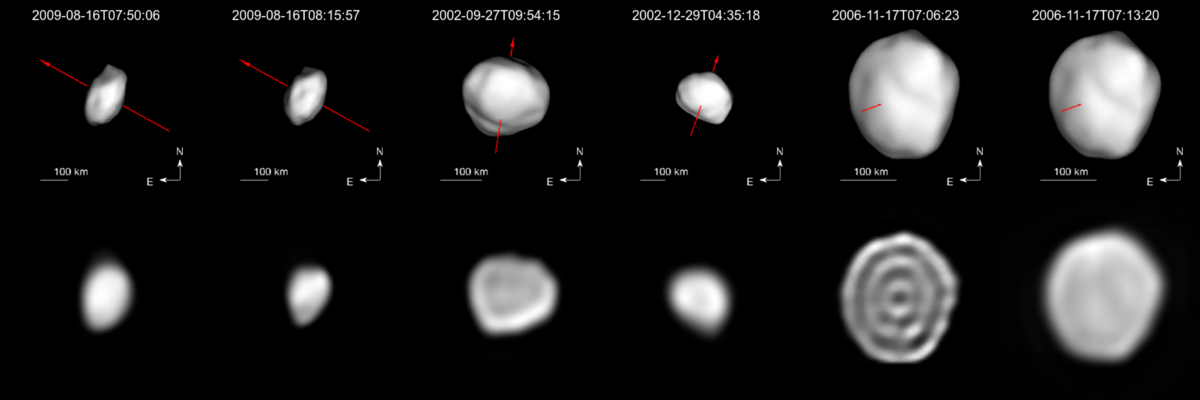}}

\resizebox{0.9\hsize}{!}{\includegraphics{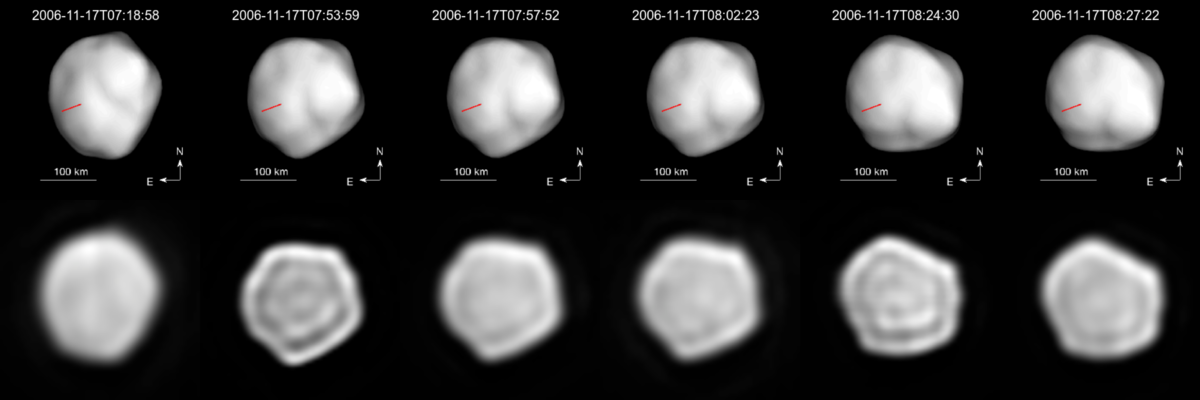}}

\resizebox{0.6\hsize}{!}{\includegraphics{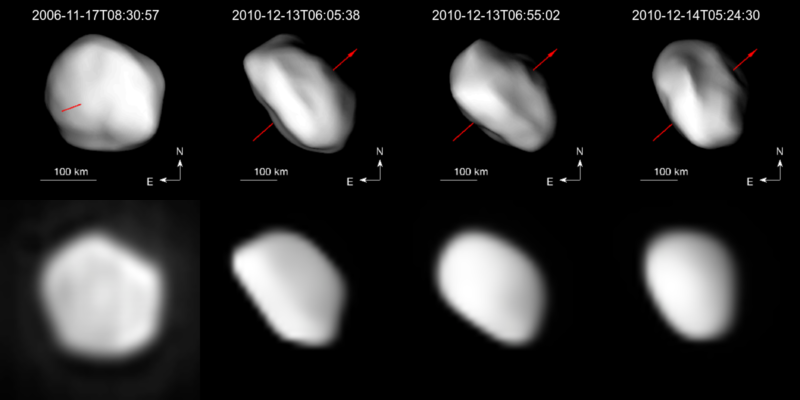}}
\caption{\label{fig:comparison2}Comparison between the Keck/NIRC2 and VLT/NaCo deconvolved images of Iris (rows 2 and 4), and the corresponding projections of our shape model (rows 1 and 3). The red line indicates the position of the rotation axis. Data affected by deconvolution artifacts were not used for the shape modeling.}
\end{figure*}
\setkeys{Gin}{draft=true}

\setkeys{Gin}{draft=false}
\begin{figure}
\begin{center}
\resizebox{\hsize}{!}{\includegraphics{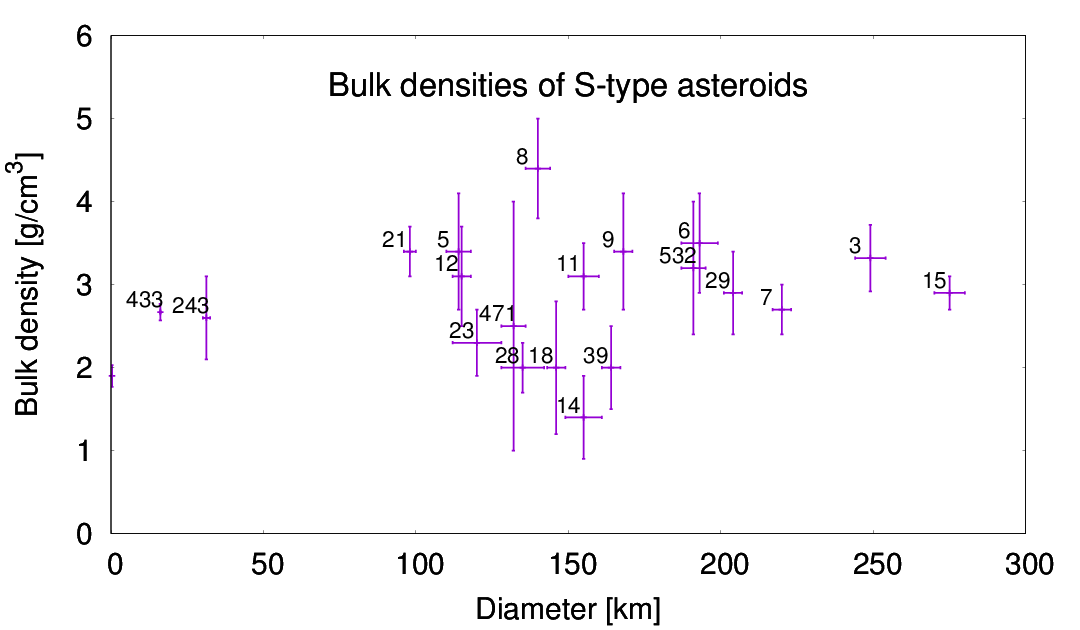}}
\end{center}
\caption{\label{fig:literature} Reliable bulk density measurements for 21 S-type asteroids compiled from the literature.}
\end{figure}
\setkeys{Gin}{draft=true}

\onecolumn
\scriptsize{
{
\begin{longtable}{r@{\,\,\,}l cc cc}
\caption{\label{tab:literature}Size and bulk density estimates of large S-type asteroids compiled from the literature. We only list  asteroids with reliable bulk density determinations. We also include our new estimates for Iris.}\\
\hline
 \multicolumn{2}{c} {Asteroid} & $D_\mathrm{V}$ & Reference & $\rho$ & Reference  \\
 \multicolumn{2}{c} {} & [km] &  & [g.cm$^{-3}$] &  \\ \hline\hline

\endfirsthead
\caption{continued.}\\

\hline
 \multicolumn{2}{c} {Asteroid} & $D_\mathrm{V}$ & Reference & $\rho$ & Reference  \\
 \multicolumn{2}{c} {} & [km] &  & [gcm$^{-3}$] &  \\ \hline\hline
\endhead
\hline
\endfoot
3   & Juno       & 249$\pm$5  & \citet{Viikinkoski2015b} & 3.32$\pm$0.40    & \citet{Viikinkoski2015b}  \\
5   & Astraea    & 114$\pm$4  & \citet{Hanus2017b}  & 3.4$\pm$0.7 & \citet{Hanus2017b} \\
6   & Hebe       & 193$\pm$6  & \citet{Marsset2017} & 3.5$\pm$0.6 & \citet{Marsset2017} \\
7   & Iris       & 214$\pm$5  & This work & 2.7$\pm$0.3 & This work \\
8   & Flora      & 140$\pm$4  & \citet{Hanus2017b} & 4.4$\pm$0.6 & \citet{Hanus2017b} \\
9   & Metis      & 168$\pm$3  & \citet{Hanus2017b} & 3.4$\pm$0.7 & \citet{Hanus2017b} \\
11  & Parthenope & 155$\pm$5  & \citet{Hanus2017b} & 3.1$\pm$0.4 & \citet{Hanus2017b} \\
12  & Victoria   & 115$\pm$3  & \citet{Viikinkoski2017} & 3.1$\pm$0.6 & \citet{Viikinkoski2017} \\
14  & Irene      & 155$\pm$6  & \citet{Viikinkoski2017} & 1.4$\pm$0.5 & \citet{Viikinkoski2017} \\
15  & Eunomia    & 275$\pm$5  & \citet{Viikinkoski2017} & 2.9$\pm$0.2 & \citet{Viikinkoski2017} \\
18  & Melpomene  & 146$\pm$3  & \citet{Hanus2017b} & 2.0$\pm$0.8 & \citet{Hanus2017b} \\
21  & Lutetia    & 98$\pm$2   & \citet{Sierks2011}  &      3.4$\pm$0.3     & \citet{Sierks2011} \\
23  & Thalia     & 120$\pm$8  & \citet{Viikinkoski2017} & 2.3$\pm$0.4 & \citet{Viikinkoski2017} \\
29  & Amphitrite & 204$\pm$3  & \citet{Hanus2017b} & 2.9$\pm$0.5 & \citet{Hanus2017b} \\
39  & Laetitia   & 164$\pm$3  & \citet{Hanus2017b} & 2.0$\pm$0.5 & \citet{Hanus2017b} \\
28  & Bellona    & 135$\pm$7  & \citet{Viikinkoski2017} & 2.0$\pm$0.3 & \citet{Viikinkoski2017} \\
243 & Ida        & 31.3$\pm$1.2 & \citet{Archinal2011}  &    2.6$\pm$0.5     & \citet{Belton1994} \\
433 & Eros       & 16.20$\pm$0.16 & \citet{Veverka2000}  &    2.67$\pm$0.10    & \citet{Veverka2000} \\
471 & Papagena   & 132$\pm$4  & \citet{Hanus2017b} & 2.5$\pm$1.5  & \citet{Hanus2017b} \\
532 & Herculina  & 191$\pm$4  & \citet{Hanus2017b} & 3.2$\pm$0.8  & \citet{Hanus2017b} \\
25143 & Itokawa  & 0.32$\pm$0.01 & \citet{Fujiwara2006}  &   1.90$\pm$0.13    & \citet{Fujiwara2006} \\ \hline
\end{longtable}
\tablefoot{
The table gives the volume-equivalent diameter $D_\mathrm{V}$ and its reference, and the bulk density $\rho$ and its reference.}
}
}

\begin{table*}
\caption{\label{tab:ao}List of disk-resolved images. For each observation the table gives the epoch, the telescope/instrument, the photometric filter, the exposure time, the airmass, the distance to the Earth $\Delta$ and the Sun $r$, the phase angle $\alpha$, the angular diameter $D_\mathrm{a}$, and the reference or the ID of the AO project.}
\centering
\begin{tabular}{rrl lrr rrr rl}
\hline 
\multicolumn{1}{c} {Date} & \multicolumn{1}{c} {UT} & \multicolumn{1}{c} {Instrument} & \multicolumn{1}{c} {Filter} & \multicolumn{1}{c} {Exp} & \multicolumn{1}{c} {Airmass} & \multicolumn{1}{c} {$\Delta$} & \multicolumn{1}{c} {$r$} & \multicolumn{1}{c} {$\alpha$} & \multicolumn{1}{c} {$D_\mathrm{a}$} & Reference or ID \\
\multicolumn{1}{c} {} & \multicolumn{1}{c} {} & \multicolumn{1}{c} {} & \multicolumn{1}{c} {} & \multicolumn{1}{c} {(s)} & \multicolumn{1}{c} {} & \multicolumn{1}{c} {(AU)} & \multicolumn{1}{c} {(AU)} & \multicolumn{1}{c} {(\degr)} & \multicolumn{1}{c} {(\arcsec)} &  \\
\hline\hline
2002-08-05  &  14:42:06  &  Keck/NIRC2  &  J         &  1.8 &  1.43  &  1.21  &  2.16  &  12.3  &  0.245  &  N10N2    \\
2002-08-05  &  14:45:29  &  Keck/NIRC2  &  H         &  1.8 &  1.45  &  1.21  &  2.16  &  12.3  &  0.245  &  N10N2    \\
2002-08-05  &  14:48:25  &  Keck/NIRC2  &  Kp        &  1.8 &  1.47  &  1.21  &  2.16  &  12.3  &  0.245  &  N10N2    \\
2002-08-05  &  15:11:44  &  Keck/NIRC2  &  Kp        &  1.8 &  1.63  &  1.21  &  2.16  &  12.3  &  0.245  &  N10N2    \\
2002-08-05  &  15:14:34  &  Keck/NIRC2  &  H         &  1.8 &  1.66  &  1.21  &  2.16  &  12.3  &  0.245  &  N10N2   \\
2002-09-27  &  09:54:15  &  Keck/NIRC2  &  Kp        &  18  &  1.29  &  1.14  &  2.03  &  17.5  &  0.260  &  \citet{Viikinkoski2017}  \\
2002-12-29  &  04:35:18  &  Keck/NIRC2  &  H         &  34  &  1.13  &  1.88  &  1.87  &  30.4  &  0.158  &  \citet{Viikinkoski2017}  \\
2005-07-17  &  07:53:59  &  Keck/NIRC2  &  Kp        &  48  &  1.37  &  2.01  &  2.78  &  16.2  &  0.147  &  \citet{Viikinkoski2017}  \\
2006-11-17  &  07:06:23  &  Keck/NIRC2  &  K         &  0.2 &  1.26  &  0.85  &  1.84  &  3.3   &  0.349  &  \citet{Viikinkoski2017}  \\
2006-11-17  &  07:13:20  &  Keck/NIRC2  &  H         &  0.2 &  1.24  &  0.85  &  1.84  &  3.3   &  0.349  &  \citet{Viikinkoski2017}  \\
2006-11-17  &  07:18:58  &  Keck/NIRC2  &  J         &  4   &  1.22  &  0.85  &  1.84  &  3.3   &  0.349  &  \citet{Viikinkoski2017}  \\
2006-11-17  &  07:53:59  &  Keck/NIRC2  &  Kp        &  2   &  1.12  &  0.85  &  1.84  &  3.3   &  0.349  &  \citet{Viikinkoski2017}  \\
2006-11-17  &  07:57:52  &  Keck/NIRC2  &  H         &  0.1 &  1.11  &  0.85  &  1.84  &  3.3   &  0.349  &  \citet{Viikinkoski2017}  \\
2006-11-17  &  08:02:23  &  Keck/NIRC2  &  J         &  2   &  1.10  &  0.85  &  1.84  &  3.3   &  0.349  &  \citet{Viikinkoski2017}  \\
2006-11-17  &  08:24:30  &  Keck/NIRC2  &  Kp        &  0.1 &  1.06  &  0.85  &  1.84  &  3.3   &  0.349  &  \citet{Viikinkoski2017}  \\
2006-11-17  &  08:27:22  &  Keck/NIRC2  &  H         &  0.1 &  1.06  &  0.85  &  1.84  &  3.3   &  0.349  &  \citet{Viikinkoski2017}  \\
2006-11-17  &  08:30:57  &  Keck/NIRC2  &  J         &  0.1 &  1.05  &  0.85  &  1.84  &  3.3   &  0.349  &  \citet{Viikinkoski2017}  \\
2009-08-16  &  07:50:06  &  Keck/NIRC2  &  PK50\_1.5 &  60  &  1.32  &  1.70  &  2.48  &  18.1  &  0.174  &  \citet{Viikinkoski2017}  \\
2009-08-16  &  08:15:57  &  Keck/NIRC2  &  PK50\_1.5 &  30  &  1.36  &  1.70  &  2.48  &  18.1  &  0.174  &  \citet{Viikinkoski2017}  \\
2010-12-13  &  06:05:38  &  VLT/NaCo    &  -         &  1   &  1.49  &  1.29  &  2.06  &  21.7  &  0.230  &  086.C-0785    \\
2010-12-13  &  06:55:02  &  VLT/NaCo    &  -         &  1   &  1.33  &  1.29  &  2.06  &  21.7  &  0.230  &  086.C-0785    \\
2010-12-14  &  05:24:30  &  VLT/NaCo    &  -         &  1   &  1.69  &  1.28  &  2.06  &  21.4  &  0.232  &  086.C-0785    \\
2017-10-10  &  3:56:12   &  VLT/SPHERE  &  N\_R &  60  &  1.74  &  0.90  &  1.85  &  13.2  &  0.329  &  199.C-0074    \\
2017-10-10  &  3:57:22   &  VLT/SPHERE  &  N\_R &  60  &  1.73  &  0.90  &  1.85  &  13.2  &  0.329  &  199.C-0074    \\
2017-10-10  &  3:58:33   &  VLT/SPHERE  &  N\_R &  60  &  1.73  &  0.90  &  1.85  &  13.2  &  0.329  &  199.C-0074    \\
2017-10-10  &  3:59:43   &  VLT/SPHERE  &  N\_R &  60  &  1.72  &  0.90  &  1.85  &  13.2  &  0.329  &  199.C-0074    \\
2017-10-10  &  4:00:55   &  VLT/SPHERE  &  N\_R &  60  &  1.72  &  0.90  &  1.85  &  13.2  &  0.329  &  199.C-0074    \\
2017-10-10  &  4:07:50   &  VLT/SPHERE  &  N\_R &  60  &  1.68  &  0.90  &  1.85  &  13.2  &  0.329  &  199.C-0074    \\
2017-10-10  &  4:09:01   &  VLT/SPHERE  &  N\_R &  60  &  1.68  &  0.90  &  1.85  &  13.2  &  0.329  &  199.C-0074    \\
2017-10-10  &  4:10:12   &  VLT/SPHERE  &  N\_R &  60  &  1.67  &  0.90  &  1.85  &  13.2  &  0.329  &  199.C-0074    \\
2017-10-10  &  4:11:22   &  VLT/SPHERE  &  N\_R &  60  &  1.67  &  0.90  &  1.85  &  13.2  &  0.329  &  199.C-0074    \\
2017-10-10  &  4:12:32   &  VLT/SPHERE  &  N\_R &  60  &  1.67  &  0.90  &  1.85  &  13.2  &  0.329  &  199.C-0074    \\
2017-10-11  &  4:40:41   &  VLT/SPHERE  &  N\_R &  60  &  1.56  &  0.89  &  1.85  &  12.7  &  0.333  &  199.C-0074    \\
2017-10-11  &  4:41:53   &  VLT/SPHERE  &  N\_R &  60  &  1.56  &  0.89  &  1.85  &  12.7  &  0.333  &  199.C-0074    \\
2017-10-11  &  4:43:05   &  VLT/SPHERE  &  N\_R &  60  &  1.56  &  0.89  &  1.85  &  12.7  &  0.333  &  199.C-0074    \\
2017-10-11  &  4:44:16   &  VLT/SPHERE  &  N\_R &  60  &  1.55  &  0.89  &  1.85  &  12.7  &  0.333  &  199.C-0074    \\
2017-10-11  &  4:45:26   &  VLT/SPHERE  &  N\_R &  60  &  1.55  &  0.89  &  1.85  &  12.7  &  0.333  &  199.C-0074    \\
2017-10-11  &  5:34:41   &  VLT/SPHERE  &  N\_R &  60  &  1.50  &  0.89  &  1.85  &  12.7  &  0.333  &  199.C-0074    \\
2017-10-11  &  5:35:52   &  VLT/SPHERE  &  N\_R &  60  &  1.50  &  0.89  &  1.85  &  12.7  &  0.333  &  199.C-0074    \\
2017-10-11  &  5:37:04   &  VLT/SPHERE  &  N\_R &  60  &  1.50  &  0.89  &  1.85  &  12.7  &  0.333  &  199.C-0074    \\
2017-10-11  &  5:38:15   &  VLT/SPHERE  &  N\_R &  60  &  1.50  &  0.89  &  1.85  &  12.7  &  0.333  &  199.C-0074    \\
2017-10-11  &  5:39:25   &  VLT/SPHERE  &  N\_R &  60  &  1.50  &  0.89  &  1.85  &  12.7  &  0.333  &  199.C-0074    \\
2017-10-11  &  6:28:33   &  VLT/SPHERE  &  N\_R &  60  &  1.54  &  0.89  &  1.85  &  12.7  &  0.333  &  199.C-0074    \\
2017-10-11  &  6:29:45   &  VLT/SPHERE  &  N\_R &  60  &  1.54  &  0.89  &  1.85  &  12.7  &  0.333  &  199.C-0074    \\
2017-10-11  &  6:30:57   &  VLT/SPHERE  &  N\_R &  60  &  1.54  &  0.89  &  1.85  &  12.7  &  0.333  &  199.C-0074    \\
2017-10-11  &  6:32:07   &  VLT/SPHERE  &  N\_R &  60  &  1.54  &  0.89  &  1.85  &  12.7  &  0.333  &  199.C-0074    \\
2017-10-11  &  6:33:18   &  VLT/SPHERE  &  N\_R &  60  &  1.54  &  0.89  &  1.85  &  12.7  &  0.333  &  199.C-0074    \\
\hline
\end{tabular}
\end{table*}

\begin{longtable}{rlr rrr l lll}
\caption{\label{tab:lcs}List of optical disk-integrated lightcurves used for \adam{} shape modeling. For each lightcurve the table gives the epoch, the number of individual measurements $N_p$, the asteroid's distances to the Earth $\Delta$ and the Sun $r$, phase angle $\varphi$, photometric filter, and observation information.}\\
\hline 
\multicolumn{1}{c} {N} & \multicolumn{1}{c} {Epoch} & \multicolumn{1}{c} {$N_p$} & \multicolumn{1}{c} {$\Delta$} & \multicolumn{1}{c} {$r$} & \multicolumn{1}{c} {$\varphi$} & \multicolumn{1}{c} {Filter} & Site & Observer  & Reference \\
 &  &  & (AU) & (AU) & (\degr) &  &  &  &  \\
\hline\hline

\endfirsthead
\caption{continued.}\\

\hline
\multicolumn{1}{c} {N} & \multicolumn{1}{c} {Epoch} & \multicolumn{1}{c} {$N_p$} & \multicolumn{1}{c} {$\Delta$} & \multicolumn{1}{c} {$r$} & \multicolumn{1}{c} {$\varphi$} & \multicolumn{1}{c} {Filter} & Site & Observer  & Reference \\
 &  &  & (AU) & (AU) & (\degr) &  &  &  &  \\
\hline\hline
\endhead
\hline
\endfoot
\hline
1   &  1950-08-12.2  &  34   &  1.70  &  2.50  &  17.4  &  V   &  MDO  & Braun, Rubingh &  \citet{Groeneveld1954}                \\
2   &  1950-08-13.2  &  25   &  1.71  &  2.50  &  17.7  &  V   &  MDO  & Braun, Rubingh &  \citet{Groeneveld1954}                \\
3   &  1950-08-14.2  &  22   &  1.72  &  2.50  &  18.0  &  V   &  MDO  & Braun, Rubingh &  \citet{Groeneveld1954}                \\
4   &  1950-08-16.2  &  9    &  1.73  &  2.49  &  18.6  &  V   &  MDO  & Braun, Rubingh &  \citet{Groeneveld1954}                \\
5   &  1952-01-28.3  &  82   &  1.18  &  2.16  &  5.2   &  V   &  MDO  & Braun, Rubingh &  \citet{Groeneveld1954}                \\
6   &  1955-12-28.5  &  39   &  1.91  &  2.38  &  23.4  &  V   &  MDO  & vH-G\&vH &  \citet{vanHouten1958}  \\
7   &  1955-12-29.4  &  39   &  1.90  &  2.38  &  23.3  &  V   &  MDO  & vH-G\&vH &  \citet{vanHouten1958}  \\
8   &  1956-01-02.5  &  35   &  1.86  &  2.39  &  22.7  &  V   &  MDO  & vH-G\&vH &  \citet{vanHouten1958}  \\
9   &  1956-01-05.5  &  18   &  1.83  &  2.40  &  22.1  &  V   &  MDO  & vH-G\&vH &  \citet{vanHouten1958}  \\
10  &  1956-03-08.4  &  64   &  1.57  &  2.55  &  4.3   &  V   &  MDO  & Kuiper &  \citet{vanHouten1958}  \\
11  &  1958-11-05.2  &  62   &  0.86  &  1.84  &  8.7   &  V   &  MDO  &  --  &  \citet{Gehrels1962}                    \\
12  &  1963-02-02.7  &  63   &  1.21  &  2.12  &  13.0  &  V   &  --  &  --  &  \citet{Chang1963}                      \\
13  &  1963-02-03.6  &  99   &  1.21  &  2.12  &  13.4  &  V   &  --  &  --  &  \citet{Chang1963}                      \\
14  &  1968-06-12.2  &  18   &  1.92  &  2.88  &  8.4   &  V   &  KPNO  &  Dunlap, Taylor  &  \citet{Taylor1977}                                \\
15  &  1968-06-13.3  &  45   &  1.93  &  2.88  &  8.8   &  V   &  KPNO  &  Dunlap, Taylor  &  \citet{Taylor1977}                                \\
16  &  1973-10-28.4  &  27   &  1.26  &  1.88  &  29.1  &  V   &  STEW  &  Dunlap  &  \citet{Taylor1977}                                \\
17  &  1973-12-15.3  &  46   &  1.00  &  1.96  &  8.8   &  V   &  KPNO  &  Taylor  &  \citet{Taylor1977}                                \\
18  &  1973-12-16.3  &  58   &  1.00  &  1.96  &  8.2   &  V   &  KPNO  &  Taylor  &  \citet{Taylor1977}                                \\
19  &  1974-02-16.3  &  17   &  1.38  &  2.10  &  22.8  &  V   &  STEW  &  Capen   &  \citet{Taylor1977}                                \\
20  &  1974-02-17.2  &  7    &  1.39  &  2.10  &  23.0  &  V   &  STEW  &  Capen   &  \citet{Taylor1977}                                \\
21  &  1980-10-14.6  &  49   &  0.97  &  1.91  &  14.3  &  V   &  --  &  --  &  \citet{Zhou1982}                            \\
22  &  1980-11-08.6  &  40   &  1.10  &  1.87  &  24.8  &  V   &  --  &  --  &  \citet{Zhou1982}                            \\
23  &  1984-09-29.4  &  52   &  1.32  &  1.84  &  31.9  &  V   &  CMC &  --  &  \citet{Lagerkvist1987b}               \\
24  &  1989-01-02.9  &  18   &  1.42  &  2.22  &  18.7  &  V   &  HLO  &  --  &  \citet{Hoffmann1993}                    \\
25  &  1989-01-04.1  &  538  &  1.42  &  2.22  &  18.4  &  V   &  HLO  &  --  &  \citet{Hoffmann1993}                    \\
26  &  1989-04-29.9  &  70   &  2.17  &  2.52  &  23.4  &  V   &  HLO  &  --  &  \citet{Hoffmann1993}                    \\
27  &  1989-05-02.9  &  42   &  2.21  &  2.52  &  23.4  &  V   &  HLO  &  --  &  \citet{Hoffmann1993}                    \\
28  &  1990-02-05.2  &  33   &  2.92  &  2.92  &  19.4  &  V   &  HLO  &  --  &  \citet{Hoffmann1993}                    \\
29  &  1990-02-06.2  &  15   &  2.90  &  2.92  &  19.5  &  V   &  HLO  &  --  &  \citet{Hoffmann1993}                    \\
30  &  1991-08-19.0  &  38   &  1.12  &  2.07  &  12.6  &  V   &  HLO  &  --  &  \citet{Hoffmann1993}                    \\
31  &  1991-09-03.0  &  75   &  1.04  &  2.04  &  6.4   &  V   &  HLO  &  --  &  \citet{Hoffmann1993}                    \\
32  &  1991-09-04.0  &  26   &  1.04  &  2.03  &  6.1   &  V   &  HLO  &  --  &  \citet{Hoffmann1993}                    \\
33  &  1991-09-05.0  &  40   &  1.03  &  2.03  &  5.9   &  V   &  HLO  &  --  &  \citet{Hoffmann1993}                    \\
34  &  1991-09-18.0  &  44   &  1.02  &  2.00  &  7.7   &  V   &  HLO  &  --  &  \citet{Hoffmann1993}                    \\
35  &  1991-11-01.9  &  9    &  1.21  &  1.92  &  26.4  &  C  &  --  &  Foglia  &  \citet{Foglia1992}                                \\
36  &  1991-11-06.9  &  23   &  1.25  &  1.91  &  27.7  &  C  &  --  &  Foglia  &  \citet{Foglia1992}                                \\
37  &  2010-12-10.1  &  623  &  1.31  &  2.05  &  22.9  &  C   &  --  &  --  &  Gerald Rousseau                            \\
38  &  2010-12-11.1  &  589  &  1.30  &  2.05  &  22.5  &  C   &  --  &  --  &  Gerald Rousseau                            \\
39  &  2013-08-15.0  &  173  &  1.18  &  2.19  &  4.4   &  C   &  --  &  --  &  Patrick Sogorb                             \\
40   &  2006-10-11    &  47   &  0.95  &  1.84  &  20.2  &  --  &  SuperWASP  &  --  &  \citet{Grice2017}                                  \\
41   &  2006-11-27    &  40   &  0.87  &  1.84  &  8.8   &  --  &  SuperWASP  &  --  &  \citet{Grice2017}                                  \\
42   &  2006-11-28    &  41   &  0.88  &  1.84  &  9.3   &  --  &  SuperWASP  &  --  &  \citet{Grice2017}                                  \\
43   &  2006-11-29    &  42   &  0.88  &  1.84  &  9.9   &  --  &  SuperWASP  &  --  &  \citet{Grice2017}                                  \\
44   &  2006-11-30    &  92   &  0.89  &  1.84  &  10.5  &  --  &  SuperWASP  &  --  &  \citet{Grice2017}                                  \\
45   &  2006-12-05    &  72   &  0.91  &  1.85  &  13.3  &  --  &  SuperWASP  &  --  &  \citet{Grice2017}                                  \\
46   &  2006-12-06    &  76   &  0.91  &  1.85  &  13.8  &  --  &  SuperWASP  &  --  &  \citet{Grice2017}                                  \\
47   &  2008-01-31    &  47   &  2.35  &  2.76  &  20.3  &  --  &  SuperWASP  &  --  &  \citet{Grice2017}                                  \\
48   &  2008-02-21    &  87   &  2.11  &  2.79  &  17.0  &  --  &  SuperWASP  &  --  &  \citet{Grice2017}                                  \\
49   &  2008-02-27    &  39   &  2.05  &  2.80  &  15.5  &  --  &  SuperWASP  &  --  &  \citet{Grice2017}                                  \\
50   &  2008-02-28    &  42   &  2.04  &  2.80  &  15.3  &  --  &  SuperWASP  &  --  &  \citet{Grice2017}                                  \\
51   &  2008-02-28    &  44   &  2.04  &  2.80  &  15.3  &  --  &  SuperWASP  &  --  &  \citet{Grice2017}                                  \\
52   &  2008-03-01    &  59   &  2.03  &  2.80  &  14.8  &  --  &  SuperWASP  &  --  &  \citet{Grice2017}                                  \\
53   &  2008-03-01    &  74   &  2.03  &  2.80  &  14.8  &  --  &  SuperWASP  &  --  &  \citet{Grice2017}                                  \\
54   &  2008-03-02    &  56   &  2.02  &  2.81  &  14.5  &  --  &  SuperWASP  &  --  &  \citet{Grice2017}                                  \\
55   &  2008-03-02    &  81   &  2.02  &  2.81  &  14.5  &  --  &  SuperWASP  &  --  &  \citet{Grice2017}                                  \\
56   &  2008-03-10    &  67   &  1.95  &  2.82  &  12.1  &  --  &  SuperWASP  &  --  &  \citet{Grice2017}                                  \\
57   &  2008-03-13    &  57   &  1.93  &  2.82  &  11.1  &  --  &  SuperWASP  &  --  &  \citet{Grice2017}                                  \\
58   &  2008-03-19    &  58   &  1.90  &  2.83  &  8.9   &  --  &  SuperWASP  &  --  &  \citet{Grice2017}                                  \\
59   &  2008-06-10    &  95   &  2.38  &  2.91  &  18.9  &  --  &  SuperWASP  &  --  &  \citet{Grice2017}                                  \\
60   &  2008-06-22    &  39   &  2.54  &  2.92  &  20.0  &  --  &  SuperWASP  &  --  &  \citet{Grice2017}                                  \\
61   &  2008-06-24    &  63   &  2.57  &  2.92  &  20.1  &  --  &  SuperWASP  &  --  &  \citet{Grice2017}                                  \\
62   &  2008-06-25    &  63   &  2.58  &  2.92  &  20.1  &  --  &  SuperWASP  &  --  &  \citet{Grice2017}                                  \\
63   &  2008-06-26    &  45   &  2.60  &  2.92  &  20.2  &  --  &  SuperWASP  &  --  &  \citet{Grice2017}                                  \\
64   &  2010-10-25    &  39   &  1.69  &  1.95  &  30.6  &  --  &  SuperWASP  &  --  &  \citet{Grice2017}                                  \\
65   &  2010-10-26    &  91   &  1.68  &  1.96  &  30.5  &  --  &  SuperWASP  &  --  &  \citet{Grice2017}                                  \\
66   &  2010-12-12    &  127  &  1.29  &  2.06  &  21.8  &  --  &  SuperWASP  &  --  &  \citet{Grice2017}                                  \\
67   &  2010-12-13    &  55   &  1.28  &  2.06  &  21.4  &  --  &  SuperWASP  &  --  &  \citet{Grice2017}                                  \\
68   &  2010-12-27    &  75   &  1.21  &  2.10  &  15.5  &  --  &  SuperWASP  &  --  &  \citet{Grice2017}                                  \\
69   &  2010-12-31    &  76   &  1.20  &  2.11  &  13.5  &  --  &  SuperWASP  &  --  &  \citet{Grice2017}                                  \\
70   &  2011-01-01    &  52   &  1.19  &  2.11  &  13.0  &  --  &  SuperWASP  &  --  &  \citet{Grice2017}                                  \\
71   &  2011-01-02    &  79   &  1.19  &  2.11  &  12.5  &  --  &  SuperWASP  &  --  &  \citet{Grice2017}                                  \\
72   &  2011-01-03    &  56   &  1.19  &  2.11  &  12.0  &  --  &  SuperWASP  &  --  &  \citet{Grice2017}                                  \\
73   &  2011-01-04    &  63   &  1.18  &  2.12  &  11.4  &  --  &  SuperWASP  &  --  &  \citet{Grice2017}                                  \\
74   &  2011-01-05    &  65   &  1.18  &  2.12  &  10.9  &  --  &  SuperWASP  &  --  &  \citet{Grice2017}                                  \\
75   &  2011-01-06    &  50   &  1.18  &  2.12  &  10.4  &  --  &  SuperWASP  &  --  &  \citet{Grice2017}                                  \\
76   &  2011-01-07    &  50   &  1.18  &  2.12  &  9.9   &  --  &  SuperWASP  &  --  &  \citet{Grice2017}                                  \\
77   &  2011-01-10    &  48   &  1.17  &  2.13  &  8.3   &  --  &  SuperWASP  &  --  &  \citet{Grice2017}                                  \\
78   &  2011-02-18    &  119  &  1.33  &  2.23  &  13.6  &  --  &  SuperWASP  &  --  &  \citet{Grice2017}                                  \\
79   &  2011-02-19    &  86   &  1.34  &  2.23  &  14.1  &  --  &  SuperWASP  &  --  &  \citet{Grice2017}                                  \\
80   &  2011-02-20    &  72   &  1.35  &  2.24  &  14.5  &  --  &  SuperWASP  &  --  &  \citet{Grice2017}                                  \\
81   &  2011-02-21    &  99   &  1.36  &  2.24  &  14.9  &  --  &  SuperWASP  &  --  &  \citet{Grice2017}                                  \\
82   &  2011-02-22    &  103  &  1.37  &  2.24  &  15.2  &  --  &  SuperWASP  &  --  &  \citet{Grice2017}                                  \\
83   &  2011-02-23    &  103  &  1.38  &  2.24  &  15.6  &  --  &  SuperWASP  &  --  &  \citet{Grice2017}                                  \\
84   &  2011-02-24    &  103  &  1.39  &  2.25  &  16.0  &  --  &  SuperWASP  &  --  &  \citet{Grice2017}                                  \\
85   &  2011-02-25    &  77   &  1.40  &  2.25  &  16.3  &  --  &  SuperWASP  &  --  &  \citet{Grice2017}                                  \\
86   &  2011-03-01    &  95   &  1.44  &  2.26  &  17.7  &  --  &  SuperWASP  &  --  &  \citet{Grice2017}                                  \\
87   &  2011-03-02    &  91   &  1.45  &  2.26  &  18.0  &  --  &  SuperWASP  &  --  &  \citet{Grice2017}                                  \\
88   &  2012-02-19    &  41   &  2.58  &  2.91  &  19.6  &  --  &  SuperWASP  &  --  &  \citet{Grice2017}                                  \\
89   &  2012-02-20    &  43   &  2.57  &  2.91  &  19.5  &  --  &  SuperWASP  &  --  &  \citet{Grice2017}                                  \\
90   &  2012-02-21    &  43   &  2.55  &  2.91  &  19.4  &  --  &  SuperWASP  &  --  &  \citet{Grice2017}                                  \\
91   &  2012-02-23    &  47   &  2.53  &  2.91  &  19.3  &  --  &  SuperWASP  &  --  &  \citet{Grice2017}                                  \\
92   &  2012-02-24    &  49   &  2.51  &  2.91  &  19.2  &  --  &  SuperWASP  &  --  &  \citet{Grice2017}                                  \\
93   &  2012-02-25    &  49   &  2.50  &  2.92  &  19.2  &  --  &  SuperWASP  &  --  &  \citet{Grice2017}                                  \\
94   &  2012-02-26    &  43   &  2.49  &  2.92  &  19.1  &  --  &  SuperWASP  &  --  &  \citet{Grice2017}                                  \\
95   &  2012-03-01    &  47   &  2.43  &  2.92  &  18.7  &  --  &  SuperWASP  &  --  &  \citet{Grice2017}                                  \\
96   &  2012-03-02    &  51   &  2.42  &  2.92  &  18.6  &  --  &  SuperWASP  &  --  &  \citet{Grice2017}                                  \\
97   &  2012-03-03    &  51   &  2.41  &  2.92  &  18.4  &  --  &  SuperWASP  &  --  &  \citet{Grice2017}                                  \\
98   &  2012-03-06    &  65   &  2.37  &  2.92  &  18.0  &  --  &  SuperWASP  &  --  &  \citet{Grice2017}                                  \\
99   &  2012-03-07    &  51   &  2.36  &  2.92  &  17.9  &  --  &  SuperWASP  &  --  &  \citet{Grice2017}                                  \\
100  &  2012-03-07    &  75   &  2.36  &  2.92  &  17.9  &  --  &  SuperWASP  &  --  &  \citet{Grice2017}                                  \\
101  &  2012-03-08    &  81   &  2.34  &  2.92  &  17.7  &  --  &  SuperWASP  &  --  &  \citet{Grice2017}                                  \\
102  &  2012-03-08    &  89   &  2.34  &  2.92  &  17.8  &  --  &  SuperWASP  &  --  &  \citet{Grice2017}                                  \\
103  &  2012-03-09    &  82   &  2.33  &  2.92  &  17.6  &  --  &  SuperWASP  &  --  &  \citet{Grice2017}                                  \\
104  &  2012-03-13    &  79   &  2.28  &  2.92  &  16.9  &  --  &  SuperWASP  &  --  &  \citet{Grice2017}                                  \\
105  &  2012-03-17    &  63   &  2.24  &  2.93  &  16.1  &  --  &  SuperWASP  &  --  &  \citet{Grice2017}                                  \\
106  &  2012-03-18    &  71   &  2.23  &  2.93  &  15.9  &  --  &  SuperWASP  &  --  &  \citet{Grice2017}                                  \\
107  &  2012-03-19    &  72   &  2.21  &  2.93  &  15.7  &  --  &  SuperWASP  &  --  &  \citet{Grice2017}                                  \\
108  &  2012-03-20    &  43   &  2.20  &  2.93  &  15.5  &  --  &  SuperWASP  &  --  &  \citet{Grice2017}                                  \\
109  &  2012-03-20    &  52   &  2.20  &  2.93  &  15.5  &  --  &  SuperWASP  &  --  &  \citet{Grice2017}                                  \\
110  &  2012-03-25    &  56   &  2.15  &  2.93  &  14.3  &  --  &  SuperWASP  &  --  &  \citet{Grice2017}                                  \\
111  &  2012-03-25    &  59   &  2.15  &  2.93  &  14.3  &  --  &  SuperWASP  &  --  &  \citet{Grice2017}                                  \\
112  &  2012-03-31    &  44   &  2.09  &  2.93  &  12.7  &  --  &  SuperWASP  &  --  &  \citet{Grice2017}                                  \\
113  &  2012-04-01    &  79   &  2.08  &  2.93  &  12.4  &  --  &  SuperWASP  &  --  &  \citet{Grice2017}                                  \\
114  &  2012-04-01    &  93   &  2.09  &  2.93  &  12.4  &  --  &  SuperWASP  &  --  &  \citet{Grice2017}                                  \\
115  &  2012-04-02    &  87   &  2.08  &  2.93  &  12.1  &  --  &  SuperWASP  &  --  &  \citet{Grice2017}                                  \\
116  &  2012-04-02    &  101  &  2.08  &  2.93  &  12.1  &  --  &  SuperWASP  &  --  &  \citet{Grice2017}                                  \\
117  &  2012-04-03    &  79   &  2.07  &  2.93  &  11.7  &  --  &  SuperWASP  &  --  &  \citet{Grice2017}                                  \\
118  &  2012-04-03    &  89   &  2.07  &  2.93  &  11.8  &  --  &  SuperWASP  &  --  &  \citet{Grice2017}                                  \\
119  &  2012-04-04    &  54   &  2.06  &  2.93  &  11.4  &  --  &  SuperWASP  &  --  &  \citet{Grice2017}                                  \\
120  &  2012-04-05    &  47   &  2.05  &  2.93  &  11.1  &  --  &  SuperWASP  &  --  &  \citet{Grice2017}                                  \\
121  &  2012-04-05    &  84   &  2.05  &  2.93  &  11.1  &  --  &  SuperWASP  &  --  &  \citet{Grice2017}                                  \\
122  &  2012-04-10    &  125  &  2.02  &  2.93  &  9.5   &  --  &  SuperWASP  &  --  &  \citet{Grice2017}                                  \\
123  &  2012-04-11    &  121  &  2.01  &  2.93  &  9.1   &  --  &  SuperWASP  &  --  &  \citet{Grice2017}                                  \\
124  &  2012-04-11    &  131  &  2.01  &  2.93  &  9.1   &  --  &  SuperWASP  &  --  &  \citet{Grice2017}                                  \\
125  &  2012-04-12    &  99   &  2.00  &  2.93  &  8.8   &  --  &  SuperWASP  &  --  &  \citet{Grice2017}                                  \\
126  &  2012-04-12    &  109  &  2.00  &  2.93  &  8.8   &  --  &  SuperWASP  &  --  &  \citet{Grice2017}                                  \\
127  &  2012-04-15    &  98   &  1.98  &  2.93  &  7.7   &  --  &  SuperWASP  &  --  &  \citet{Grice2017}                                  \\
128  &  2012-04-20    &  55   &  1.96  &  2.94  &  5.8   &  --  &  SuperWASP  &  --  &  \citet{Grice2017}                                  \\
129  &  2012-04-23    &  43   &  1.95  &  2.94  &  4.7   &  --  &  SuperWASP  &  --  &  \citet{Grice2017}                                  \\
130  &  2012-04-29    &  55   &  1.93  &  2.94  &  2.6   &  --  &  SuperWASP  &  --  &  \citet{Grice2017}                                  \\
131  &  2012-04-29    &  57   &  1.93  &  2.94  &  2.6   &  --  &  SuperWASP  &  --  &  \citet{Grice2017}                                  \\
132  &  2012-04-29    &  59   &  1.93  &  2.94  &  2.6   &  --  &  SuperWASP  &  --  &  \citet{Grice2017}                                  \\
133  &  2012-06-06    &  58   &  2.07  &  2.93  &  12.6  &  --  &  SuperWASP  &  --  &  \citet{Grice2017}                                  \\
\hline
\end{longtable}
\tablefoot{
     MDO -- McDonald Observatory, KPNO -- Kitt Peak National Observatory, 91cm or 41 telescopes. STEW -- Steward 51-cm Observatory, CMC --Carlsberg Meridian Circle at La Palma, HLO -- Hoher List Observatory, vH-G\&vH -- van Houten-Groeneveld, van Houten.
    }

\begin{table*}
\begin{center}
  \caption[Mass estimates of (7) Iris]{\label{tab:mass}
    Mass estimates ($\mathcal{M}$) of (7)~Iris from the literature.
    For each, the 1\,$\sigma$ uncertainty, method, and 
    bibliographic reference are listed. The methods are
    \textsc{defl}: Deflection, \textsc{ephem}: Ephemeris.
  }
  \begin{tabular}{rrll}
    \hline\hline
     \multicolumn{1}{c}{\#} & \multicolumn{1}{c}{Mass ($\mathcal{M}$)} &
     \multicolumn{1}{c}{Method} & \multicolumn{1}{c}{Reference}  \\
    & \multicolumn{1}{c}{(kg)} \\
    \hline
  1 & $(3.98 \pm 1.79) \times 10^{19}$                   & \textsc{defl}   & \citet{1999-IAA-Vasiliev}                \\
  2 & $(11.90 \pm 1.99) \times 10^{18}$                  & \textsc{defl}   & \citet{2001-IAA-Krasinsky}               \\
  3 & $(2.80 \pm 0.28) \times 10^{19}$                   & \textsc{defl}   & \citet{2002-ACM-Chernetenko}             \\
  4 & $(2.80 \pm 0.28) \times 10^{19}$                   & \textsc{defl}   & \citet{2004-SoSyR-38-Kochetova}          \\
  5 & $(10.30 \pm 1.59) \times 10^{18}$                  & \textsc{defl}   & \citet{2004-COSPAR-35-Pitjeva}           \\
  6 & $(1.25 \pm 0.06) \times 10^{19}$                   & \textsc{ephem}  & \citet{2005-SoSyR-39-Pitjeva}            \\
  7 & $(1.79 \pm 0.20) \times 10^{19}$                   & \textsc{ephem}  & \citet{2007-ASPC-370-Aslan}              \\
  8 & $(1.36 \pm 0.10) \times 10^{19}$                   & \textsc{defl}   & \citet{2008-DPS-40-Baer}                 \\
  9 & $(4.77 \pm 0.60) \times 10^{19}$                   & \textsc{defl}   & \citet{2008-PSS-56-Ivantsov}             \\
 10 & $(1.15 \pm 0.02) \times 10^{19}$                   & \textsc{ephem}  & \citet{2008-AA-477-Fienga}               \\
 11 & $(11.90 \pm 1.29) \times 10^{18}$                  & \textsc{ephem}  & \citet{2009-SciNote-Folkner}             \\
 12 & $(6.56 \pm 1.59) \times 10^{18}$                   & \textsc{ephem}  & \citet{2010-IAUS-261-Pitjeva}            \\
 13 & $(1.62 \pm 0.09) \times 10^{19}$                   & \textsc{defl}   & \citet{2011-AJ-141-Baer}                 \\
 14 & $(11.00 \pm 2.63) \times 10^{18}$                  & \textsc{ephem}  & \citet{2011-Icarus-211-Konopliv}         \\
 15 & $(1.75 \pm 0.29) \times 10^{19}$                   & \textsc{defl}   & \citet{2011-AJ-142-Zielenbach}           \\
 16 & $(1.72 \pm 0.16) \times 10^{19}$                   & \textsc{defl}   & \citet{2011-AJ-142-Zielenbach}           \\
 17 & $(1.68 \pm 0.16) \times 10^{19}$                   & \textsc{defl}   & \citet{2011-AJ-142-Zielenbach}           \\
 18 & $(2.33 \pm 0.31) \times 10^{19}$                   & \textsc{defl}   & \citet{2011-AJ-142-Zielenbach}           \\
 19 & $(11.30 \pm 0.80) \times 10^{18}$                  & \textsc{ephem}  & \citet{2011-DPS-Fienga}                  \\
 20 & $(12.50 \pm 1.21) \times 10^{18}$                  & \textsc{ephem}  & \citet{2012-SciNote-Fienga}              \\
 21 & $(14.80 \pm 1.65) \times 10^{18}$                  & \textsc{ephem}  & \citet{2013-Icarus-222-Kuchynka}         \\
 22 & $(1.30 \pm 0.06) \times 10^{19}$                   & \textsc{ephem}  & \citet{2013-SoSyR-47-Pitjeva}            \\
 23 & $(11.60 \pm 0.97) \times 10^{18}$                  & \textsc{ephem}  & \citet{2014-SciNote-Fienga}              \\
 24 & $(1.39 \pm 0.04) \times 10^{19}$                   & \textsc{defl}   & \citet{2014-AA-565-Goffin}               \\
 25 & $(1.39 \pm 0.06) \times 10^{19}$                   & \textsc{defl}   & \citet{2014-SoSyR-48-Kochetova}          \\
 26 & $(10.10 \pm 0.56) \times 10^{18}$                  & \textsc{ephem}  & \citet{2017-BDL-108-Viswanathan}         \\
 27 & $4.18_{-1.39}^{+4.62} \times 10^{18}$             & \textsc{defl}   & \citet{2017-Icarus-297-Siltala}          \\
 28 & $(1.65 \pm 0.09) \times 10^{19}$                   & \textsc{ephem}  & \citet{2017-AJ-154-Baer}                 \\
 29 & $(1.67 \pm 0.12) \times 10^{19}$                   & \textsc{ephem}  & \citet{2017-AJ-154-Baer}                 \\
 30 & $(7.24 \pm 0.57) \times 10^{18}$                   & \textsc{ephem}  & \citet{2018-INPOP-Fienga}                \\
\hline
 & \textbf{$(13.75 \pm 1.30) \times 10^{18}$} & \multicolumn{2}{c}{Median (1$\sigma$ uncertainty)} \\
   \hline
  \end{tabular}
\end{center}
\end{table*}

\end{appendix}
\end{document}